\documentclass[fleqn,usenatbib]{mnras}

\usepackage{savesym}
\usepackage{txfonts}
\savesymbol{iint}
\savesymbol{iiint}
\savesymbol{iiiint}
\savesymbol{idotsint}

\usepackage[T1]{fontenc}
\usepackage{ae,aecompl}

\usepackage{graphicx}	% Including figure files
\usepackage{subfig}
\usepackage{amsmath}	% Advanced maths commands
\restoresymbol{AMS}{iint}
\restoresymbol{AMS}{iiint}
\restoresymbol{AMS}{iiiint}
\restoresymbol{AMS}{idotsint}
\usepackage{amssymb}	% Extra maths symbols
\usepackage[shortlabels]{enumitem}
\usepackage{xcolor}

\definecolor{darkgreen}{RGB}{0,128,0}

\newcommand{\norm}[1]{\lVert #1 \rVert}

\title[Ultra-deep TDEs]{Ultra-deep tidal disruption events: prompt self-intersections and observables}

\author[S. Darbha et al.]{
Siva Darbha,$^{1}$\thanks{E-mail: siva.darbha@berkeley.edu}
Eric R.~Coughlin,$^{2,4}$\thanks{Einstein fellow}
Daniel Kasen$^{1,2,3}$
and Chris Nixon$^{5}$
\\
$^{1}$Department of Physics, University of California, Berkeley, Berkeley, CA 94720, USA\\
$^{2}$Department of Astronomy and Theoretical Astrophysics Center, University of California, Berkeley, Berkeley, CA 94720, USA\\
$^{3}$Nuclear Science Division, Lawrence Berkeley National Laboratory, Berkeley, CA 94720, USA\\
$^{4}$Columbia Astrophysics Laboratory, Columbia University, New York, NY, 10027, USA\\
$^{5}$Theoretical Astrophysics Group, Department of Physics and Astronomy, University of Leicester, Leicester LE1 7RH, UK
}

\date{Accepted XXX. Received YYY; in original form ZZZ}

\pubyear{2019}

\begin{document}
\label{firstpage}
\pagerange{\pageref{firstpage}--\pageref{lastpage}}
\maketitle

\begin{abstract}

\noindent
A star approaching a supermassive black hole (SMBH) can be torn apart in a tidal disruption event (TDE). We examine ultra-deep TDEs, a new regime in which the disrupted debris approaches close to the black hole's Schwarzschild radius, and the leading part intersects the trailing part at the first pericenter passage. We calculate the range of penetration factors $\beta$ vs SMBH masses $M$ that produce these prompt self-intersections using a Newtonian analytic estimate and a general relativistic (GR) geodesic model. We find that significant self-intersection of Solar-type stars requires $\beta \sim 50 - 127$ for $M/M_\odot = 10^4$, down to $\beta \sim 5.6 - 5.9$ for $M/M_\odot = 10^6$. We run smoothed-particle hydrodynamic (SPH) simulations to corroborate our calculations and find close agreement, with a slightly shallower dependence on $M$. We predict that the shock from the collision emits an X-ray flare lasting $t \sim 2$ s with $L \sim 10^{47}$ ergs/s at $E \sim 2$ keV, and the debris has a prompt accretion episode lasting $t \sim$ several min. The events are rare and occur with a rate $\dot{N} \lesssim 10^{-7}$ Mpc$^{-3}$ yr$^{-1}$. Ultra-deep TDEs can probe the strong gravity and demographics of low-mass SMBHs.

\end{abstract}

% Select between one and six entries from the list of approved keywords. Don't make up new ones.

%%%%%%%%%%%%%%%%%%%%%%%%%%%%%%%%%%%%%%%%%%%%%%%%%%

%%%%%%%%%%%%%%%%% BODY OF PAPER %%%%%%%%%%%%%%%%%%

\section{Introduction}
\label{sec:intro}

Stars can be destroyed by the gravitational field of a supermassive black hole (SMBH) if they reach its tidal radius $r_t \simeq R_* (M/M_*)^{1/3}$, where the SMBH has mass $M$ and the star has mass $M_*$ and radius $R_*$. Within the tidal radius, the tidal gravity of the SMBH exceeds the self-gravity of the star, and the star is stretched into a stream of debris \citep{kochanek94, coughlin16b}. This process of stellar destruction by a black hole is known as a tidal disruption event (TDE).

The qualitative timeline of a canonical TDE likely proceeds as follows. If the disrupted star originates near or beyond the sphere of influence of the SMBH, which constitutes the most likely radius from which Solar-like stars are scattered into the loss cone (e.g., \citealt{stone16}), then the star's center of mass (CM) is effectively on a parabolic orbit. At disruption, the stellar debris acquires a range of binding energies due to the tidal potential of the SMBH \citep{lacy82}, which binds half of the debris to to the black hole and unbinds the other half \citep{rees88}. The event is likely ``dark'' in this early phase, characterized by the radial expansion of the stream with relatively little emission, though there are some potential signatures \citep{carter82,kobayashi04,guillochon09,kasen10,yalinewich19}. When the debris returns to pericenter, general relativistic (GR) apsidal precession causes the debris apocenter to deviate from its Keplerian value. This deflection causes the outgoing and incoming material to intersect, dissipate kinetic energy through shocks, and form an accretion disk after several orbits \citep{cannizzo90,kochanek94,lee96,kim99,hayasaki13,hayasaki16b,guillochon14,bonnerot16,shiokawa15}. When the black hole mass satisfies $M \lesssim 10^{7}M_{\odot}$, the fallback rate is super-Eddington (for full disruptions of Solar-like stars; \citealt{evans89,wu18}), which can lead to the production of radiation driven winds \citep{strubbe09} and jets \citep{coughlin14}. The accretion (and associated outflows) produces a highly luminous, short-lived emission event, in which the lightcurve rapidly rises, reaches a peak, and decays as a power-law.

While this timeline is likely upheld for most TDEs, the detailed appearance of a given TDE depends on the properties of the disrupted star, the SMBH, and the specifics of the stellar orbit (e.g. the pericenter). The event horizon of a non-rotating black hole is determined by the Schwarzschild radius, $r_S = 2r_g$, where $r_g = GM/c^2$ is the gravitational radius. The encounter strength can be parametrized by the penetration factor $\beta = L_t^2 / L_\textrm{cm}^2$, where $L_t^2 = 2GM r_t$ and $L_\textrm{cm}$ is the specific angular momentum of the stellar CM. In Newtonian encounters, the stellar CM angular momentum is $L_\textrm{cm}^2 = 2GM r_p$, where $r_p$ is the CM pericenter distance, and the penetration factor becomes $\beta = r_t / r_p$ \citep{carter82}. We can also define $r_p$ in this way even for relativistic encounters, though in this case it is simply a parameter and has diminished physical significance; the true GR pericenter $(r_p)_{GR}$ will be different here, and will even be undefined if the star is captured by the SMBH.

The general picture of a canonical TDE given above has been confirmed numerically for large tidal radii ($r_t \gg r_S$) and shallow encounters ($\beta \sim 1$, with $r_p \sim r_t$). Simulations using smoothed-particle hydrodynamics (SPH) and adaptive-mesh grid-based methods have observed the predicted spread in energies (for early work, see \citealt{evans89}; for more recent work, see \citealt{lodato09,guillochon13,coughlin15}), and simulations using SPH and magnetohydrodynamics (MHD) have observed the stream-stream collisions leading to debris circularization \citep{bonnerot16,hayasaki16b,jiang16}. Analytic estimates \citep{lodato11} and detailed radiative transfer calculations \citep{strubbe15,roth16,roth18} have supported the expected behavior of the lightcurve and characterized its spectral features.

However, the situation can change dramatically for large tidal radii ($r_t \gg r_S$) when the star undergoes an ultra-deep encounter ($\beta \gg 1$, with $(r_p)_{GR} \sim r_S$). In this case, the star is tidally stretched for a long duration as it approaches the BH; indeed, when its CM reaches $(r_p)_{GR}$, the star's leading and trailing edges can be displaced much farther apart than the initial stellar radius $R_*$. In addition, since the apsidal precession angle $\Delta \phi \gtrsim 2\pi$ when $(r_p)_{GR} \sim r_S$, the leading edge of the extremely stretched star can conceivably intersect the trailing edge \textit{before} the latter reaches its pericenter. In this case, the stream will intersect itself roughly at pericenter, the debris will shock heat and dissipate orbital energy, and some fraction of the star will accrete onto the SMBH almost instantaneously at this first pericenter passage. These extreme TDEs will thus exhibit no delay between disruption and accretion and, consequently, show a very different rise-peak-decay signature. This distinct signature may allow us to probe the tidal deformation of the star near pericenter and the debris dynamics in strong gravity.

There have been efforts to examine the signatures of deep encounters of main sequence stars on SMBHs \citep{bicknell83,laguna93,brassart08,brassart10,evans15,sadowski16,tejeda17}, though not in the ultra-deep regime that produces prompt self-intersections. The results still reveal interesting features that may also occur in our case. For instance, \citet{evans15} examine disruptions of main sequence stars in deep encounters ($\beta = 10, 15$) with SMBHs of mass $M = 10^5 M_\odot$. Their simulations exhibit an early accretion burst followed by a flat accretion rate at later times; strong GR effects modify the $\dot{M} \sim t^{-5/3}$ late-time accretion rate expected of canonical TDEs.

There has been greater focus on deep encounters of white dwarfs (WDs) on stellar and intermediate-mass BHs \citep{luminet89b,frolov94,rosswog09,haas12,macleod16,tanikawa17,kawana18,anninos18}, since these are more promising as sites of nuclear ignition near pericenter and as sources for gravitational wave (GW) emission. As in the case of main sequence stars, much of this work has not directly examined the prompt self-intersection regime. One notable exception is \citet{kawana18}, who study deep encounters using 3D SPH simulations coupled with a nuclear reaction $\alpha$-network, and observe prompt self-intersections at pericenter for encounters with a WD of mass $M_{WD} = 0.6 M_\odot$, a SMBH of mass $M = 10 M_\odot$, and penetration factor $\beta = 5$. They label these ``Type III TDEs,'' and find that these energetic collisions heat the debris and efficiently circularize it, but the heating is not sufficient to ignite nuclear reactions; they note, though, that their simulations likely do not fully resolve the collision numerically.

In this paper we investigate the possibility of prompt self-intersections from ultra-deep TDEs of main sequence stars. In Section \ref{sec:models}, we outline our parameter regime and present two models to calculate the range of encounter depths for which we expect prompt self-intersections to occur; first, we derive a simple, order-of-magnitude estimate in Newtonian gravity (Section \ref{subsec:analytic}), and second, we model the geodesics of debris elements in the Schwarzschild metric under the impulse (or ``frozen-in'') approximation (Section \ref{subsec:geodesic}). In Section \ref{sec:simulations}, we use SPH simulations to corroborate our estimated range of encounter depths for which prompt self-intersections occur. In Section \ref{sec:discussion}, we conclude and discuss the possible signatures from these extreme TDEs.

\section{Models}
\label{sec:models}

We defined the tidal radius $r_t$, the penetration factor $\beta$, and the Schwarzschild radius $r_S$ in Section \ref{sec:intro}. These parametrize the nature of the encounter. We use the Newtonian expression for $r_t$, which differs from the relativistic expression only marginally for the parameters that we examine below \citep{kesden12a,servin17}. We note that some authors adopt the definition $\beta = r_t / (r_p)_{GR}$ when studying relativistic TDEs (e.g. \citealt{guillochon13,tejeda17}); this is different from our definition above, so one must exercise care when making comparisons.

We examine TDEs whose parameters satisfy the hierarchy $R_*, r_S \ll r_t$. The condition $r_S \ll r_t$ permits deep ($\beta \gg 1$) encounters, and can be expressed as
\begin{equation}
\frac{r_S}{r_t} \simeq 9.1 \times 10^{-3} M_5^{2/3} m_*^{1/3} r_*^{-1} \ll 1
\label{eq:rsoverrt}
\end{equation}
where $m_* \equiv M_* / M_\odot$, $r_* \equiv R_* / R_\odot$, and $M_5 \equiv M / (10^5 M_\odot)$. The condition $R_* \ll r_t$ allows us to use first order expressions for the spread in coordinates, energy, and angular momentum across the star, and can be expressed as
\begin{equation}
\frac{R_*}{r_t} \simeq 2.2 \times 10^{-2} M_5^{-1/3} m_*^{1/3} \ll 1
\label{eq:rstaroverrt}
\end{equation}
We restrict our attention to ratios $\lesssim 5 \times 10^{-2}$. For solar-type stars, this implies SMBH masses in the range $M \sim (10^4 - 10^6) M_\odot$. For stars of mass $M_* / M_\odot = 10$ and radius $R_* = R_\odot (M_* / M_\odot)^\alpha$ with $\alpha = 0.57$, which holds for main sequence stars with masses $M_* / M_\odot \gtrsim 1$ \citep{torres10}, this implies SMBH masses in the range $M \sim (5 \times 10^4 - 5 \times 10^6) M_\odot$.

For highly penetrating encounters ($\beta \gg 1$), the stellar debris experiences strong tidal compression near pericenter, which leads to a rapid pressure increase and the generation of a shock during the bounce phase \citep{carter82,carter83,bicknell83}. \citet{carter82} used geometrical arguments and the adiabatic approximation to estimate that, for a $\gamma = 5/3$ polytrope, 
%the compression velocity is roughly $u \simeq \beta (GM_* / R_*)^{1/2}$, and 
the central stellar density $\rho_*$ and temperature $T_*$ are compressed to maximum values of $\rho_m \simeq \beta^3 \rho_*$ and $T_m \simeq \beta^2 T_*$ near pericenter, and argued that this would ignite nuclear reactions. This process has been explored in subsequent work using detailed hydrodynamic simulations, and there remains uncertainty over the extent of compression, the plausibility of nuclear reactions, and the effects on the debris orbits approaching pericenter \citep{carter83,bicknell83,laguna93,kobayashi04,guillochon09,brassart08,brassart10,stone13,evans15,tejeda17}. Most recently, \citet{tejeda17} ran extensive SPH simulations for TDEs by Kerr BHs and found strong compression in encounters with $\beta \gtrsim 10$; \citet{evans15}, in contrast, performed GR hydrodynamic simulations, and did not observe strong compression in encounters with $\beta \sim 10 - 15$ (for solar-type stars and $M = 10^5 M_\odot$ SMBHs), and instead found that the star has a large spatial spread before pericenter passage due to tidal stretching in the orbital plane.

In this section, we seek simple estimates for the range of $\beta$ for which the disrupted debris promptly self-intersects. We thus ignore hydrodynamic effects at pericenter, namely strong compression, shock generation, and nuclear detonation. While these effects could alter the qualitative behavior of the debris following self-intersection, and thus the appearance of the event, they should not drastically impact the ability of the star to promptly self-intersect in the first place.

\subsection{Analytic Model}
\label{subsec:analytic}

In this subsection, we derive an order-of-magnitude expression in Newtonian gravity for the values of $\beta$ for which we expect the star to undergo self-intersection prior to leaving pericenter.

Prior to reaching pericenter, we can effectively model the stellar CM on a purely radial orbit since its pericenter is very small relative to the tidal radius. For a parabolic orbit, the energy of the CM is $\epsilon = 0$. If the take the CM to be at $r_t$ at time $t=0$, then we can write its position as an explicit function of time as
\begin{equation}
r_\textrm{cm} = r_t \left( 1 - \frac{3}{2} \tilde{t} \right)^{2/3},
\end{equation}
where $\tilde{t} = t / \sqrt{r_t^3 / 2GM}$. The CM thus reaches its pericenter $r_p$ at time
\begin{equation}
\tilde{t}_p = \frac{2}{3} (1 - \beta^{-3/2}).
\end{equation}
The tidal field 
%The Newtonian tidal tensor acts to 
stretches the star in the radial direction and compresses it in the two orthogonal directions. If $R$ is the radial extent of the stretched star, then for $R \ll r_\textrm{cm}$, the tidal acceleration in the radial direction is
\begin{equation}
\ddot{R} = \left(\frac{r_t}{r_\textrm{cm}}\right)^3 R = R \left( 1 - \frac{3}{2} \tilde{t} \right)^{-2}
\end{equation}
where the overdot denotes differentiation with respect to $\tilde{t}$. With the initial conditions $R(0) = 2R_*$ (the stellar radius is $R_*$ at the tidal radius) and $\dot{R}(0) = 0$ (the star is in hydrostatic equilibrium at the tidal radius) the solution for the radial extent of the star as a function of time is
\begin{equation}
R = R_* \left[ \frac{8}{5} \left( 1 - \frac{3}{2} \tilde{t} \right)^{-1/3} + \frac{2}{5} \left( 1 - \frac{3}{2} \tilde{t} \right)^{4/3} \right].
\end{equation}
By the time the CM reaches $r_p$, the condition $R \ll r_\textrm{cm}$ is no longer satisfied; nevertheless, if we apply our result to obtain a rough estimate, then the star at this point has a radial extent of
\begin{equation}
R_p = R_* \left( \frac{8}{5} \beta^{1/2} + \frac{2}{5} \beta^{-2} \right) \simeq \frac{8}{5} R_* \beta^{1/2},
\end{equation}
where the last line follows from the assumption $\beta \gg 1$. We note that the leading order scaling $R_p \sim \beta^{1/2}$ is equal to that found in Eq. B6 of \citet{stone13} for the long principal axis of a tidally deformed star on a parabolic orbit. The higher order terms differ, though, likely because we evaluate the deformation at different points along the orbit.

As the stretched star passes through pericenter, it will become long enough to intersect itself if $R_p \geq 2\pi r_p$; using the expressions for $\beta$ and $r_t$, this becomes
\begin{equation}
\beta \geq \beta_c \simeq \left(\frac{5\pi}{4}\right)^{2/3} \left(\frac{M}{M_*}\right)^{2/9} \simeq 32 M_5^{2/9} m_*^{-2/9}
\label{eq:betaoomestimate}
\end{equation}
which is independent of the stellar radius and only depends weakly on the mass of the SMBH.

Equation \eqref{eq:betaoomestimate} does not explicitly incorporate GR effects. In particular, 1) some or all of the star may be captured by the BH, 2) the relative precession angle of uncaptured debris must be $\Delta \phi > 2\pi$ for a self-intersection to occur, and 3) the GR tidal deformation is stronger (in a static, spherically symmetric spacetime) than the Newtonian tidal field at a given radial coordinate \citep{luminet85,servin17}. Nevertheless, stars will become stretched appreciably as above only in ultra-deep encounters, for which we also expect large apsidal precession, and thus this expression provides a rough, order-of-magnitude estimate of the $\beta$ required to achieve a prompt self-intersection.

\subsection{Geodesic Model}
\label{subsec:geodesic}

In this subsection, we model the disrupted debris on independent Schwarzschild geodesics to estimate the range of encounter depths for which it promptly intersects itself. Here and for the remainder of this paper, we use geometric units $G = c = 1$ and a metric signature ($-,+,+,+$). We often express tensors by their symbols alone. We also use early Latin indices $a,b,\hdots$ to label tensors in abstract index notation, and Greek indices to labels components. In addition, we use Greek indices to refer to spacetime components $\mu=0,1,2,3$, and middle Latin indices $i,j,\hdots$ to refer to spatial components $i=1,2,3$.

We integrate the geodesics using the Runge-Kutta-Fehlberg 78 integrator from the C++ \textsc{boost} libraries, setting the absolute and relative errors to $\epsilon_\textrm{abs} = 10^{-12}$ and $\epsilon_\textrm{rel} = 10^{-10}$. This is a high-order adaptive integrator that has been used to accurately model null and timelike geodesics for general relativistic ray tracing in strong gravity \citep{vincent11,grould16}.

%The relativistic tidal tensor for a spherically symmetric 
The tidal force at large radii (compared to the gravitational radius) varies as $\sim 1/r^3$, where $r$ is the radial coordinate from the SMBH. The strength of the SMBH's tidal force compared to a star's gravitational self-force thus increases rapidly as the star crosses $r_t$, which suggests that the tidal force can be modeled as an impulsive effect that is activated at $r_t$. Specifically, for $r \gtrsim r_t$, tidal effects are negligible and we can model the star as if it retains perfect hydrostatic balance; for $r \lesssim r_t$, tidal effects are dominant and we can model the motion of stellar gas parcels as independent orbits in the gravitational field of the SMBH. The orbital elements (e.g. specific energy and angular momentum) of a given gas parcel are therefore ``frozen in'' once the stellar center of mass reaches the tidal radius \citep{lodato09}.

Recently, \citet{steinberg19} showed that the ``frozen-in'' approximation does not adequately capture the energy distribution of the debris, which is modified by self-gravity throughout the encounter, though the gas parcels are still well-modeled by individual orbits, a result validated by our simulations (Section \ref{sec:simulations}). In addition, the star likely experiences strong tidal compression and shock generation at pericenter, and thus the ``frozen-in'' approximation is not strictly valid at this point. The star also experiences additional tidal deformations prior to reaching the tidal radius \citep{lodato09,coughlin15}. It is sufficient, though, for our estimates in this subsection to ignore these difficulties. We will address the importance of the latter in Section \ref{sec:simulations} with the aid of hydrodynamic simulations. The assumption of geodesic orbits should also be particularly well-maintained for the equatorial plane of the star, about which the tidal compression is symmetric.

The gravity of a non-spinning and charge-less SMBH of mass $M$ is described by the Schwarzschild metric, which is the general exterior solution for a static, spherically symmetric spacetime \citep{birkhoff23,hawking73}. The metric $g$ has the associated line element \citep{chandrasekhar83}
\begin{equation}
ds^2 = -\left( 1 - \frac{2M}{r} \right) dt^2 + \left( 1 - \frac{2M}{r} \right)^{-1} dr^2 + r^2 d\theta^2 + r^2 \sin^2\theta d\phi^2
\end{equation}
where we express it in Schwarzschild coordinates $x^\mu = (x^0, x^1, x^2, x^3) = (t,r,\theta,\phi)$. A timelike geodesic $\gamma$ has 4-velocity $u = \gamma'(\tau) = \dot{x}^\mu \partial_\mu$, where we parametrize the geodesic by the proper time $\tau$ and where the overdot denotes differentiation with respect to $\tau$, $\dot{x}^\mu \equiv \frac{d(x^\mu \circ \gamma)}{d\tau}$. The 4-velocity satisfies $g_{\mu \nu} \dot{x}^\mu \dot{x}^\nu = -1$. The spatial motion of a geodesic is restricted to a 2D plane, which we can set equal to the equatorial plane ($\theta = \pi/2$, $\dot{\theta} = 0$) due to spherical symmetry. 

% The metric admits the Killing vectors $\partial_t$ and $\partial_\phi$, leading to the conserved quantities $E = -g(u,\partial_t)$ (the specific energy) and $L = g(u,\partial_\phi)$ (the specific angular momentum). The equations of motion of the geodesic components are
% \begin{align}
% \dot{t} &= \left( 1 - \frac{2M}{r} \right)^{-1} E \\
% \label{eq:phidot}
% \dot{\phi} &= \frac{L}{r^2} \\
% \dot{r}^2 &= E^2 - 2V(r) \\
% V(r) &= \frac{1}{2} - \frac{M}{r} + \frac{L^2}{2r^2} - \frac{ML^2}{r^3}
% \end{align}
% where $V(r)$ is the 1D effective potential.

%A $1$-parameter collection of geodesics experiences relative acceleration determined by the local curvature. If the collection has the tangent vector field $T$ and deviation vector field $S$, then the relative acceleration of nearby geodesics is given by the geodesic deviation equation
%\begin{equation}
%T^c \nabla_c ( T^b \nabla_b S^a ) = R^a_{\;\:bcd} T^b T^c S^d
%\end{equation}
%where $R^a_{\;\:bcd}$ is the Riemann curvature tensor, which acts as $R^a_{\;\:bcd} Z^b = \nabla_c \nabla_d Z^a - \nabla_d \nabla_c Z^a$ for a vector field $Z^a$.

The Lagrangian for a free particle in geodesic motion is $\mathcal{L} = \frac{1}{2} g_{\mu \nu} \dot{x}^\mu \dot{x}^\nu$, and the corresponding Hamiltonian is $\mathcal{H} = p_\mu \dot{x}^\mu - \mathcal{L} = \frac{1}{2} g^{\mu\nu} p_\mu p_\nu$, where $p_\mu = \frac{\partial \mathcal{L}}{\partial \dot{x}^\mu}$ are the (covariant) canonical momenta. The equations of motion can be obtained from Hamilton's equations, $\dot{x}^\mu = \frac{\partial \mathcal{H}}{\partial p_\mu}$ and $\dot{p}_\mu = -\frac{\partial \mathcal{H}}{\partial x^\mu}$. The Hamiltonian is independent of $t$ and $\phi$, so $\dot{p}_t = \dot{p}_\phi = 0$, leading to the conserved quantities $E = -p_t$ (the specific energy) and $L = p_\phi$ (the specific angular momentum). The equations of motion can thus be written as
\begin{align}
\dot{t} &= \left( 1 - \frac{2M}{r} \right)^{-1} E \\
\label{eq:phidot}
\dot{\phi} &= \frac{L}{r^2} \\
\dot{r} &= \left( 1 - \frac{2M}{r} \right) p_r \\
\dot{p}_r &= -\frac{M}{r^2} p^2_r -\frac{M E^2}{(r-2M)^2} + \frac{L^2}{r^3}
\end{align}
The radial momentum can be written explicitly as
\begin{align}
p_r^2 &= \left( 1 - \frac{2M}{r} \right)^{-2} \left[ E^2 - 2V(r) \right] \\
V(r) &= \frac{1}{2} - \frac{M}{r} + \frac{L^2}{2r^2} - \frac{ML^2}{r^3}
\end{align}
where $V(r)$ is the 1D effective potential.

Let $\alpha$ be the CM geodesic parametrized by $\tau$, with 4-velocity $u_\textrm{cm} = \alpha'(\tau) = \dot{x}_\textrm{cm}^\mu \partial_\mu$. The CM has angular momentum $L_\textrm{cm}$, which must satisfy $L_\textrm{cm} \lesssim 2Mr_t$ for a disruption to occur, and energy $E_\textrm{cm} = 1$, since the incident unbound stars approach the SMBH on roughly parabolic orbits (though these can change under different physical circumstances; \citealt{stone11,coughlin17,darbha18}). The stellar gas parcels around the CM have a range of coordinates, energies, and angular momenta at the moment of disruption. These can be calculated rigorously and self-consistently in terms of Fermi Normal Coordinates (FNCs) defined along $\alpha$ \citep{manasse63}. We construct the FNCs using the general approach of previous work \citep{luminet85,brassart10,kesden12b}, which we briefly summarize.

Let $\lambda_{(\mu)}$ be an orthonormal tetrad (ONT) defined along $\alpha$, where the circular brackets label the tetrad elements. The timelike vector $\lambda_{(0)}$ is equal to the tangent of $\alpha$ (i.e. the 4-velocity) and the tetrad is parallel-propagated along $\alpha(\tau)$. All together,
\begin{align}
g(\lambda_{(\mu)},\lambda_{(\nu)}) &= \eta_{(\mu)(\nu)} \\
\lambda_{(0)} &= u_\textrm{cm} \\
\nabla_{\lambda_{(0)}} \lambda_{(\mu)} &= 0
\end{align}
where $\eta$ is the Minkowski metric. The indices in circular brackets are raised and lowered using $\eta$, and those not in circular brackets are done so using $g$ \citep{chandrasekhar83,wald84}. We use the tetrad given by \citet{luminet85} (corrected in \citealt{brassart10}), and present them in Appendix \ref{sec:appendixa}. We note that these were obtained from the more general expressions calculated by \citet{marck83} in the Kerr metric.

Consider a debris element at a point $q$ in the neighborhood of $\alpha$. We take $\alpha(0) = p_0$ as an arbitrary reference point. There is a unique spacelike geodesic $\chi$ that passes through $q$ and is orthogonal to $\alpha$ at some point $\alpha(\tau)=p$, where we parametrize $\chi$ by the proper distance $s$. We take $\chi(0)=p$, so $\chi(s)=q$ for some $s$. Let $X = X^{(i)} \lambda_{(i)}$ be the spacelike vector that is tangent to $\chi$ at $p$ such that the proper distance to $q$ along $\chi$ is $s = \norm{X}$, where $\norm{X} = [g(X,X)]^{1/2} = [X^{(i)} X_{(i)}]^{1/2}$. The FNCs of the debris element at $q$ are defined to be $(\tau, X^{(i)})$.

We take the reference point $p_0$ to be the point at which the CM is at the tidal radius. We note that $X$ is parallel-propagated along $\alpha$ by construction, and so the stellar gas parcels do not accelerate relative to the CM geodesic before disruption. The gas parcels fall in the range $\norm{X} \leq R_*$. For convenience, we define the unscaled vector $\tilde{X}$ by $X = R_* \tilde{X}$, where $\norm{\tilde{X}} \leq 1$.

%The geodesic deviation equation for $X$ can be written in the tetrad basis as
%\begin{equation}
%\frac{d^2 X^{(i)}}{d\tau^2} = C^{(i)}_{\;\;\;\,(j)} X^{(j)}
%\end{equation}
%where $C_{(i)(j)} = - R_{abcd} (\lambda_{(0)})^a (\lambda_{(i)})^b (\lambda_{(0)})^c (\lambda_{(j)})^d$ is the relativistic tidal tensor. The tidal tensor components were calculated in previous work \citep{luminet85,brassart10}, and we present them in Appendix \ref{sec:appendixa}. 

To first order, the debris element with FNCs $(0, X^{(i)})$ has Schwarzschild coordinates $x^\mu = x_\textrm{cm}^\mu + X^{(i)} \lambda_{(i)}^\mu$, and energy and angular momentum
\begin{align}
\label{eq:EandL1}
E(X^{(i)}) &= E_\textrm{cm} + \Delta E (X^{(i)}) \\
L(X^{(i)}) &= L_\textrm{cm} + \Delta L (X^{(i)})
\end{align}
with deviations given by \citep{kesden12b}
\begin{align}
\Delta E(X^{(i)}) &= \nabla_X E_\textrm{cm} = -g_{\mu\nu} \lambda^\mu_{(0)} X^{(i)} \lambda_{(i)}^\beta \Gamma^\nu_{\beta t} \\
\Delta L(X^{(i)}) &= \nabla_X L_\textrm{cm} = g_{\mu\nu} \lambda^\mu_{(0)} X^{(i)} \lambda_{(i)}^\beta \Gamma^\nu_{\beta \phi}
\end{align}
where $\nabla$ is the Levi-Civita connection and $\Gamma$ are the Christoffel symbols. We can write these as
\begin{align}
\Delta E(X^{(i)}) &= \frac{M}{r_\textrm{cm}^2} \sum_{i=1,3} X^{(i)} (\dot{t}_\textrm{cm} \lambda_{(i)}^r - \dot{r}_\textrm{cm} \lambda_{(i)}^t) \\
\label{eq:deltaEanddeltaL2}
\Delta L(X^{(i)}) &= r_\textrm{cm} \sum_{i=1,3} X^{(i)} (-\dot{r}_\textrm{cm} \lambda_{(i)}^\phi + \dot{\phi}_\textrm{cm} \lambda_{(i)}^r)
\end{align}
We note that these do not depend on $X^{(2)}$ explicitly, only through the constraint $\norm{X} \leq R_*$. We only need the first order expressions due to the parameter regime that we consider (Eqs. \ref{eq:rsoverrt} - \ref{eq:rstaroverrt}), though the extension to a more general regime would require higher order corrections \citep{ishii05,cheng13,cheng14}.

In our setup, we set the stellar CM to move in the equatorial plane ($\theta=\pi/2$, $\dot{\theta}=0$) and to be at the tidal radius along the $x$-axis $(r,\theta,\phi) = (r_t,\pi/2,0)$. We restrict our attention to the post-disruption debris geodesics also in the equatorial plane ($X^{(2)} = 0$). This is the symmetry plane around which the debris moves. 
% ; for $E_\textrm{cm} = 1$ and $\beta \gg 1$, the entire star is confined to an angle $\alpha \simeq R_* / (r_p r_t)^{1/2}$ centered on this plane \citep{carter82}.
This restriction to equatorial debris reduces the degrees of freedom to the $(t,r,\phi)$ coordinates, which simplifies the condition for self-intersection. The equatorial geodesics are uniquely identified by the coordinates $(X^{(1)},X^{(3)})$, which map in a linear, one-to-one fashion to the parameters $(E,L)$ for a given $E_\textrm{cm}$, $L_\textrm{cm}$, and $r_t$.

The ultimate fate of a debris element with parameters $(E,L)$ can be determined from $V(r)$. If $L^2 < 12M^2$, the debris will undoubtedly be captured by the SMBH. If $L^2 \geq 12M^2$, the debris will be captured (including stalling indefinitely) if $V(r_-) \leq \frac{E^2}{2}$, where $r_-(L) \equiv \frac{L^2}{2M} \left[ 1 - \left( 1 - \frac{12M^2}{L^2} \right)^{1/2} \right]$ is the radius of the unstable circular orbit, and the angular momentum for capture $L_\textrm{cap}$ is defined from $V(r_-(L_\textrm{cap})) = \frac{E^2}{2}$. In particular, geodesics with $E=1$ will be captured if $L^2 \leq L_\textrm{cap}^2 = 16M^2$.

\begin{figure*}
\subfloat[$\tilde{X}^{(1)} \tilde{X}^{(3)}$-plane]{\includegraphics[scale=0.34]{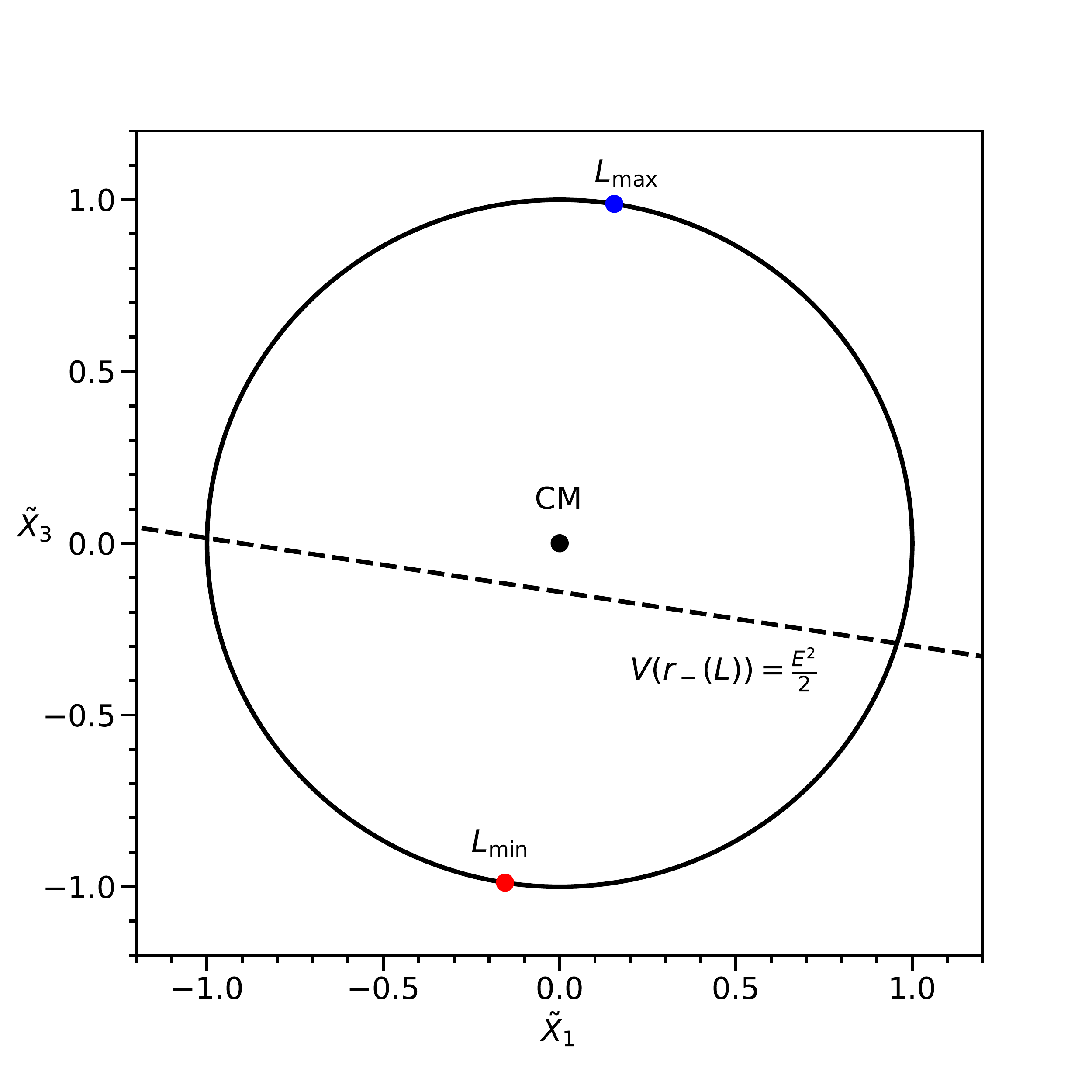}}\hfill
\subfloat[$EL$-plane]{\includegraphics[scale=0.34]{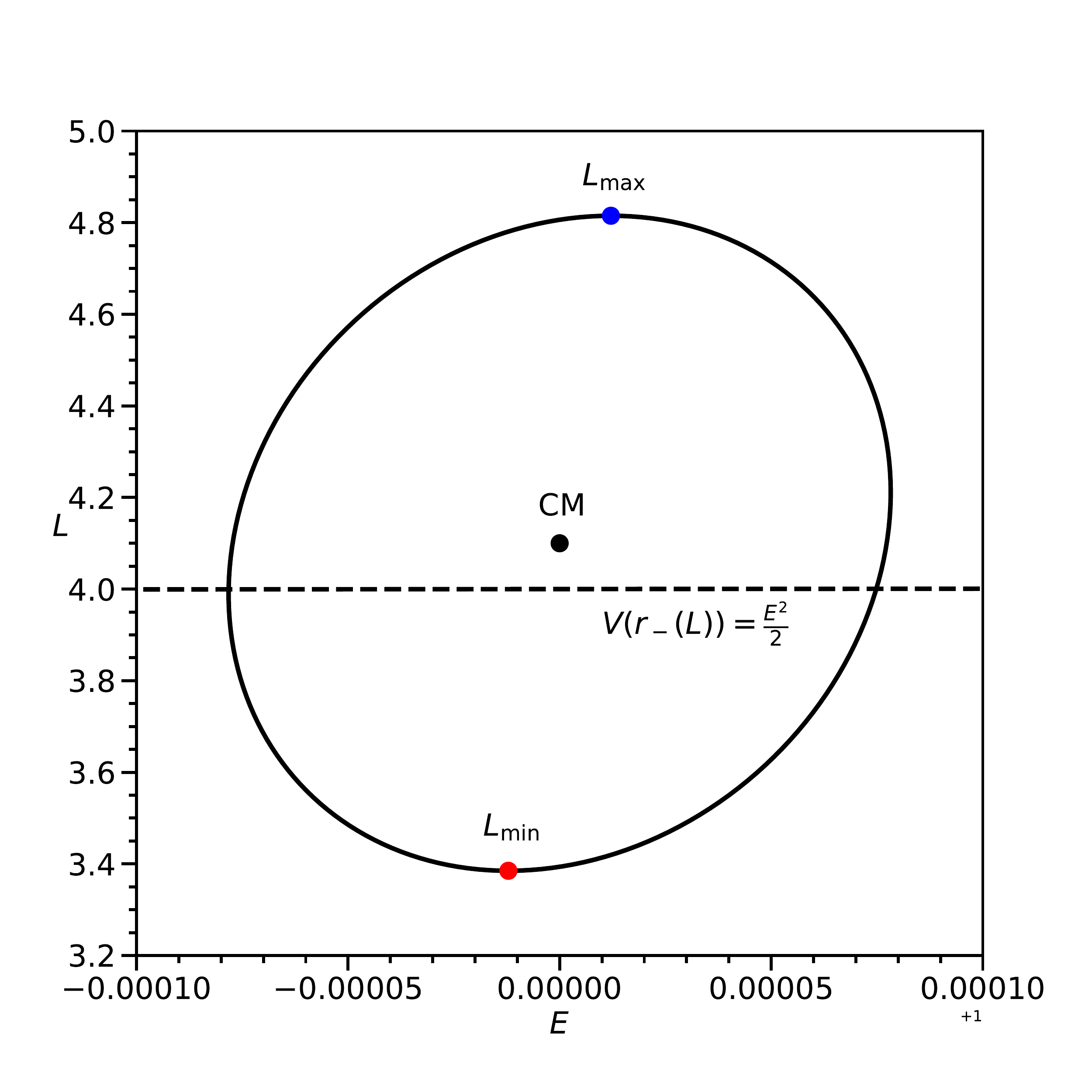}}
\caption{The debris in the equatorial plane ($X^{(2)} = 0$) for $M/M_\odot = 5 \times 10^4$, solar-type stars, $E_\textrm{cm} = 1$, and $L_\textrm{cm} = 4.1 M$, described in terms of a) the coordinates ($\tilde{X}^{(1)}$, $\tilde{X}^{(3)}$) and b) the parameters ($E$,$L$). The black dashed line shows the curve separating captured (below) and uncaptured (above) debris elements. The black circle marks the CM. The red (blue) dot shows the geodesic with the smallest (largest) value of $L$.}
\label{fig:debrisdomains}
\end{figure*}

Figure \ref{fig:debrisdomains} shows the coordinates and parameters of the equatorial debris, and the regions which are captured and uncaptured, for one set of SMBH and stellar parameters. The geometry of the regions can be understood simply. The debris elements in the equatorial plane have coordinates in the domain $\norm{\tilde{X}}^2 = \tilde{X}^{(i)} \tilde{X}_{(i)} = (\tilde{X}^{(1)})^2 + (\tilde{X}^{(3)})^2 \leq 1$. The linear map in Eqs. \ref{eq:EandL1} - \ref{eq:deltaEanddeltaL2} maps this circle in the $\tilde{X}^{(1)} \tilde{X}^{(3)}$-plane to the ellipse $A(E-E_\textrm{cm})^2 + B(E-E_\textrm{cm})[(L-L_\textrm{cm})/M] + C[(L-L_\textrm{cm})/M]^2 \leq 1$ in the $EL$-plane, where $A,B,C$ each depend on $E_\textrm{cm}$, $L_\textrm{cm}$, $r_\textrm{cm}$. Since $E_\textrm{cm} = 1$ and $\Delta E / E_\textrm{cm} \ll 1$, the curve $V(r_-) = E^2/2$ in the $EL$-plane that separates the captured and uncaptured debris is roughly a horizontal line at $L = L_\textrm{cap} = 4M$, and is mapped in inverse to the line shown in the $\tilde{X}^{(1)} \tilde{X}^{(3)}$-plane.

The debris domains in Figure \ref{fig:debrisdomains} do not take stellar structure into account. In particular, the geodesic model treats all points in these domains equally, whereas in any physical star the density profile is peaked at the center and decays towards $R_*$. This difference becomes particularly important for $L_\textrm{cm}$ in the range $L_\textrm{cm} < L_\textrm{cap} < L_\textrm{cm} + (\Delta L)_\textrm{max}$. In the geodesic model, the CM will be captured in this range, but there will be uncaptured debris that may produce a self-intersection. However, in a physical star, this uncaptured debris will have low density since it arises from the stellar envelope, and will thus only weakly self-intersect.

To express the prompt self-intersection condition for the equatorial debris, let $\psi \left(u,\tau\right)$ be the collection of equatorial debris geodesics, $\psi: \mathbb{R}^2 \times \mathbb{R} \to \mathcal{M}$, where $u = (\tilde{X}^{(1)}, \tilde{X}^{(3)}) \in \mathbb{R}^2$ labels a geodesic in the collection by its (unscaled) spatial FNCs before disruption, $\tau \in \mathbb{R}$ gives the proper time along a geodesic, and $\mathcal{M}$ is the spacetime manifold. A self-intersection will occur if there exist two distinct points $u_1$ and $u_2$ and some times $\tau_1$ and $\tau_2$ such that $\left(x^\mu \circ \psi\right) \left(u_1,\tau_1\right) = \left(x^\mu \circ \psi\right) \left(u_2,\tau_2\right)$, where equivalence in the $\phi$-coordinate is defined such that the two expressions differ by at least $2\pi$.

\begin{figure*}
\subfloat[]{\includegraphics[scale=0.34]{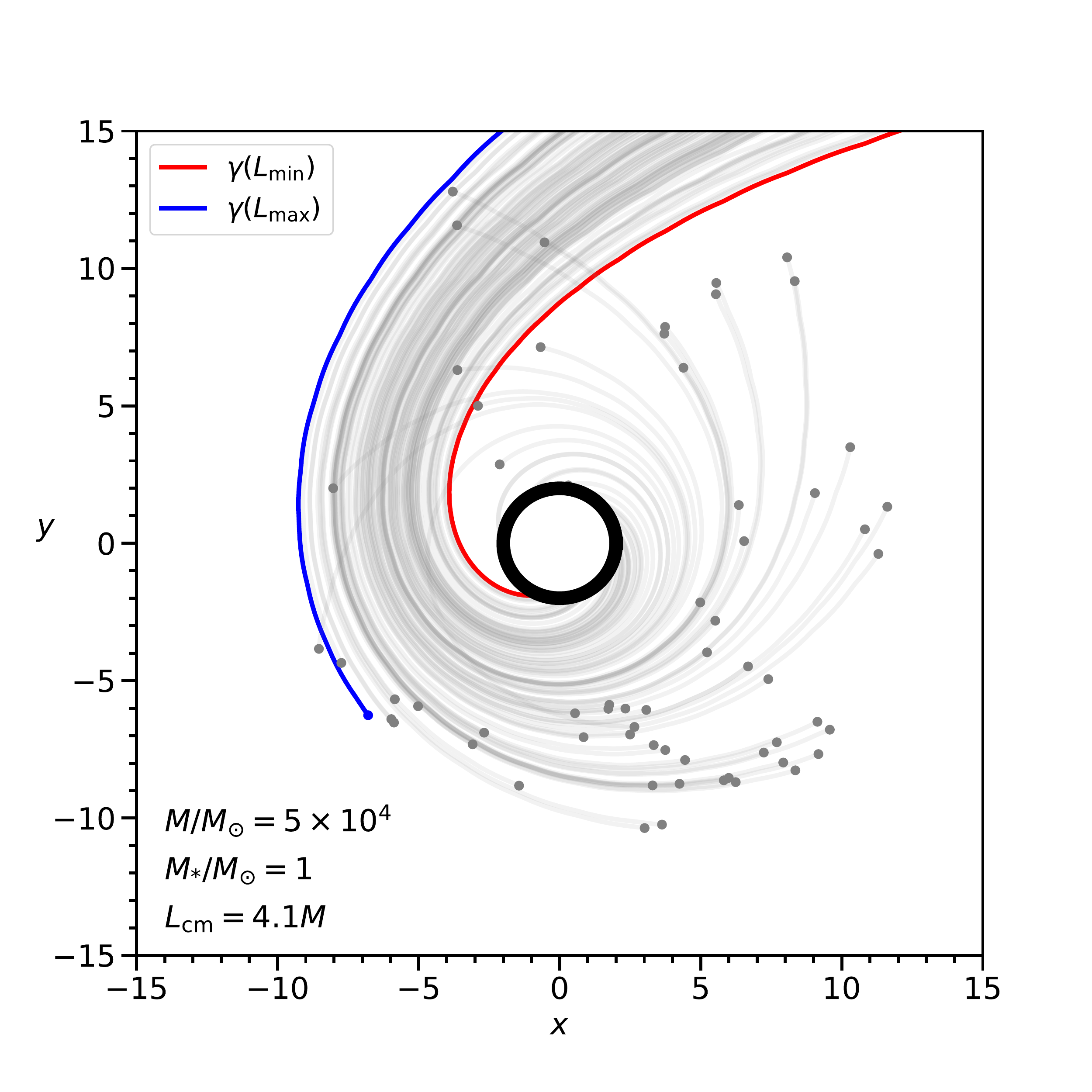}}\hfill
\subfloat[]{\includegraphics[scale=0.34]{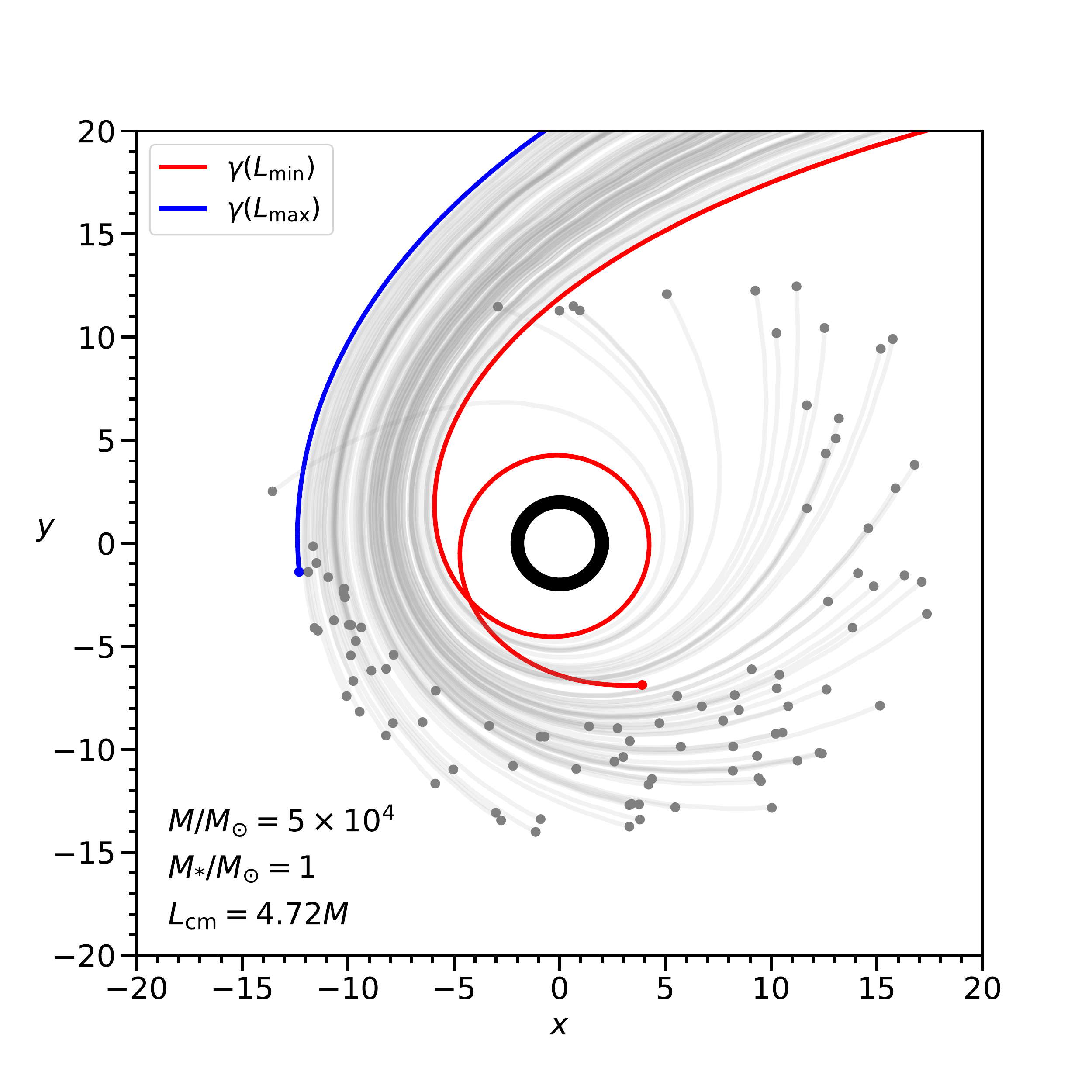}}
\caption{Orbit snapshots at a coordinate time $t$ showing the self-intersection behavior of the equatorial debris geodesics, for $M/M_\odot = 5 \times 10^4$, solar-type stars, $E_\textrm{cm} = 1$, and two different values of $L_\textrm{cm}$. The black circle shows the Schwarzschild radius $r_S = 2M$, the location of the event horizon. The red (blue) curve shows the geodesic with the smallest (largest) value of $L$. The grey curves show $100$ geodesics with randomly selected values of $E$ and $L$. a) $L_\textrm{cm} = 4.1M$. This falls in the regime $V(r_-(L_\textrm{max})) > E(L_\textrm{max})$ and $V(r_-(L_\textrm{min})) \leq E(L_\textrm{min})$, in which some of the geodesics are captured and some of the uncaptured geodesics must promptly self-intersect. b) $L_\textrm{cm} = 4.72M$. This falls in the regime $V(r_-(L_\textrm{min})) > E(L_\textrm{min})$, in which none of the geodesics are captured. Here, $\Delta \phi \equiv \phi(L_\textrm{min}) - \phi(L_\textrm{max})  > 2\pi$ for some $t$, so the debris promptly self-intersects.
}
\label{fig:geodesics}
\end{figure*}

However, this mathematical relation is cumbersome to implement numerically, so we instead adopt an approximate condition for self-intersection. If $V(r_-(L_\textrm{max})) \leq E(L_\textrm{max})$, then all of the debris will be captured. If $V(r_-(L_\textrm{max})) > E(L_\textrm{max})$ and $V(r_-(L_\textrm{min})) \leq E(L_\textrm{min})$, then some (but not all) of the debris will be captured, and some of the uncaptured debris must promptly self-intersect. This is because the $EL$ parameter space is smooth, so there will be geodesics with pericenters arbitrarily close to the event horizon that will rapidly precess and intersect the geodesics that have larger pericenters. If $V(r_-(L_\textrm{min})) > E(L_\textrm{min})$, then none of the debris will be captured. In this case, we select the debris elements with the smallest and largest values of $L$, which we label as $\gamma(L_\textrm{min})$ and $\gamma(L_\textrm{max})$. These two have among the smallest and largest values of $(r_p)_{GR}$, and will be among the leading and trailing geodesics to reach the SMBH. We integrate these geodesics, and classify the outcome as a prompt self-intersection if there is a coordinate time $t$ at which $\Delta \phi \equiv \phi(L_\textrm{min}) - \phi(L_\textrm{max}) \geq 2\pi$. If this condition is met, then there likely exist two geodesics that satisfy the exact condition above. We note that this approximate condition is neither necessary nor sufficient for the exact condition. If $\left(x^\mu \circ \psi\right) \left(u_1,\tau_1\right) = \left(x^\mu \circ \psi\right) \left(u_2,\tau_2\right)$, then it is possible that $\Delta \phi < 2\pi$ since $\gamma(L_\textrm{min})$ and $\gamma(L_\textrm{max})$ are not rigorously the leading and trailing geodesics. If $\Delta \phi \geq 2\pi$, then it is possible that $\left(x^\mu \circ \psi\right) \left(u_1,\tau_1\right) \neq \left(x^\mu \circ \psi\right) \left(u_2,\tau_2\right)$ because $\gamma(L_\textrm{min})$ fails to catch up to any of the other geodesics. These two possibilities, though, are unlikely. Figure \ref{fig:geodesics} shows the trajectories of geodesics for one set of SMBH and stellar parameters and two different values of $L_\textrm{cm}$, which produce prompt self-intersections in these two regimes.

\begin{figure*}
\subfloat[]{\includegraphics[scale=0.34]{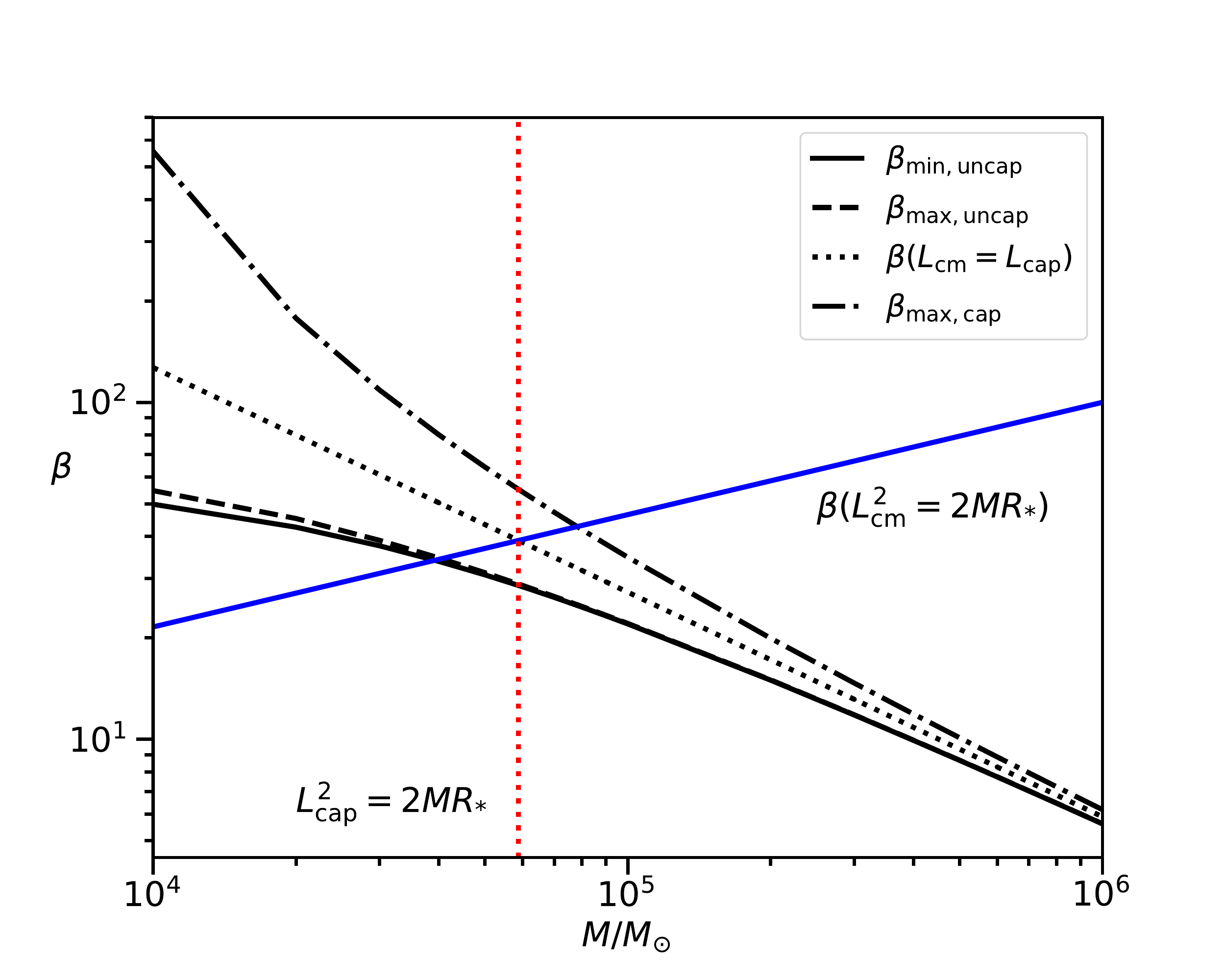}}\hfill
\subfloat[]{\includegraphics[scale=0.34]{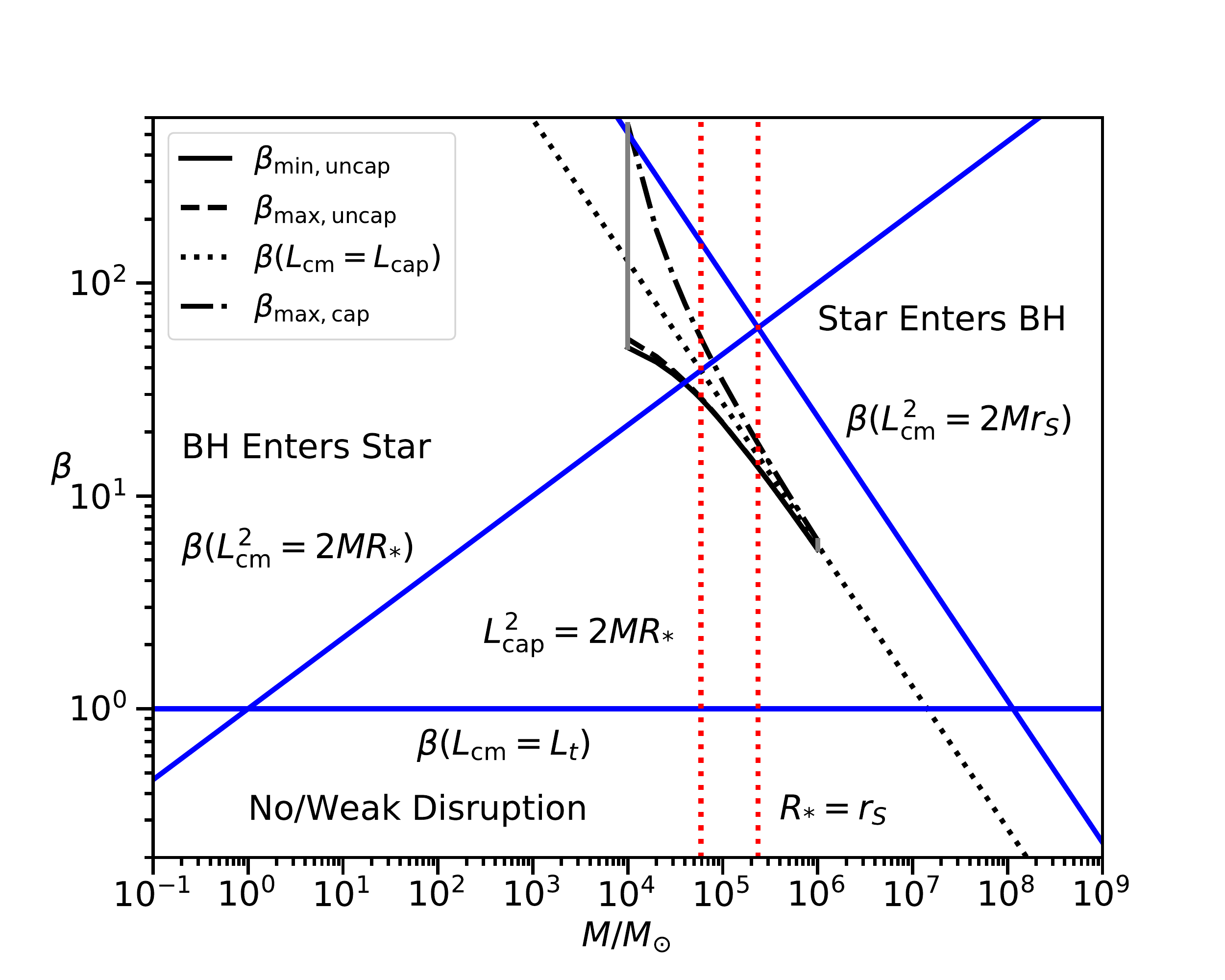}}
\caption{The range of $\beta = L_t^2 / L_\textrm{cm}^2$ over which the debris will promptly self-intersect (black curves). The solid (dashed) curve shows the minimum (maximum) value $\beta_\textrm{min,uncap}$ ($\beta_\textrm{max,uncap}$) for which there is a self intersection and in which none of the debris is captured. The dotted curve shows the GR boundary for the star's CM to get captured by the BH, $\beta(L_\textrm{cm}=L_\textrm{cap})$. The dot-dashed curve shows the value $\beta_\textrm{max,cap}$ at which all of the debris gets captured. The blue, solid lines show the Newtonian boundaries for the star to absorb the BH, $\beta(L_\textrm{cm}^2=2MR_*)$, the star's CM to get captured by the BH, $\beta(L_\textrm{cm}^2=2Mr_s)$, and the star to avoid disruption, $\beta(L_\textrm{cm}=L_t)$. The red, dotted, vertical lines show the SMBH masses at which $L^2_\textrm{cap} = 2MR_*$ and $R_* = r_S$. The disrupted stars are Sun-like ($M_* / M_\odot = 1$, $R_* / R_\odot = 1$). a) $\beta$ vs $M / M_\odot$ for the range of BH masses for which our approximations hold. b) The ``TDE triangle'' (adapted from \citealt{luminet89}). The grey, solid, vertical lines show the boundary of the mass range in panel a).}
\label{fig:betamstarovermsol1}
\end{figure*}

\begin{figure*}
\subfloat[]{\includegraphics[scale=0.34]{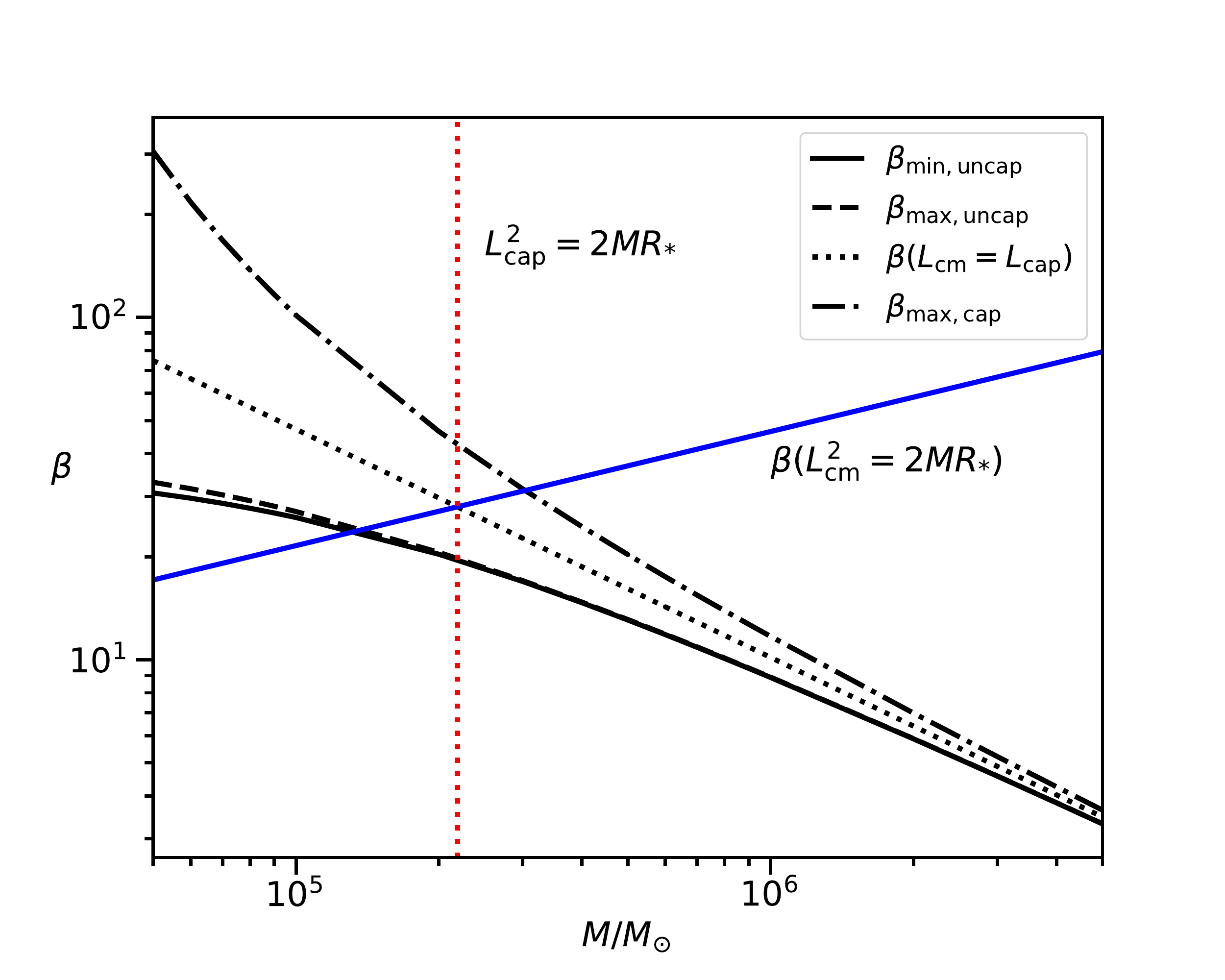}}\hfill
\subfloat[]{\includegraphics[scale=0.34]{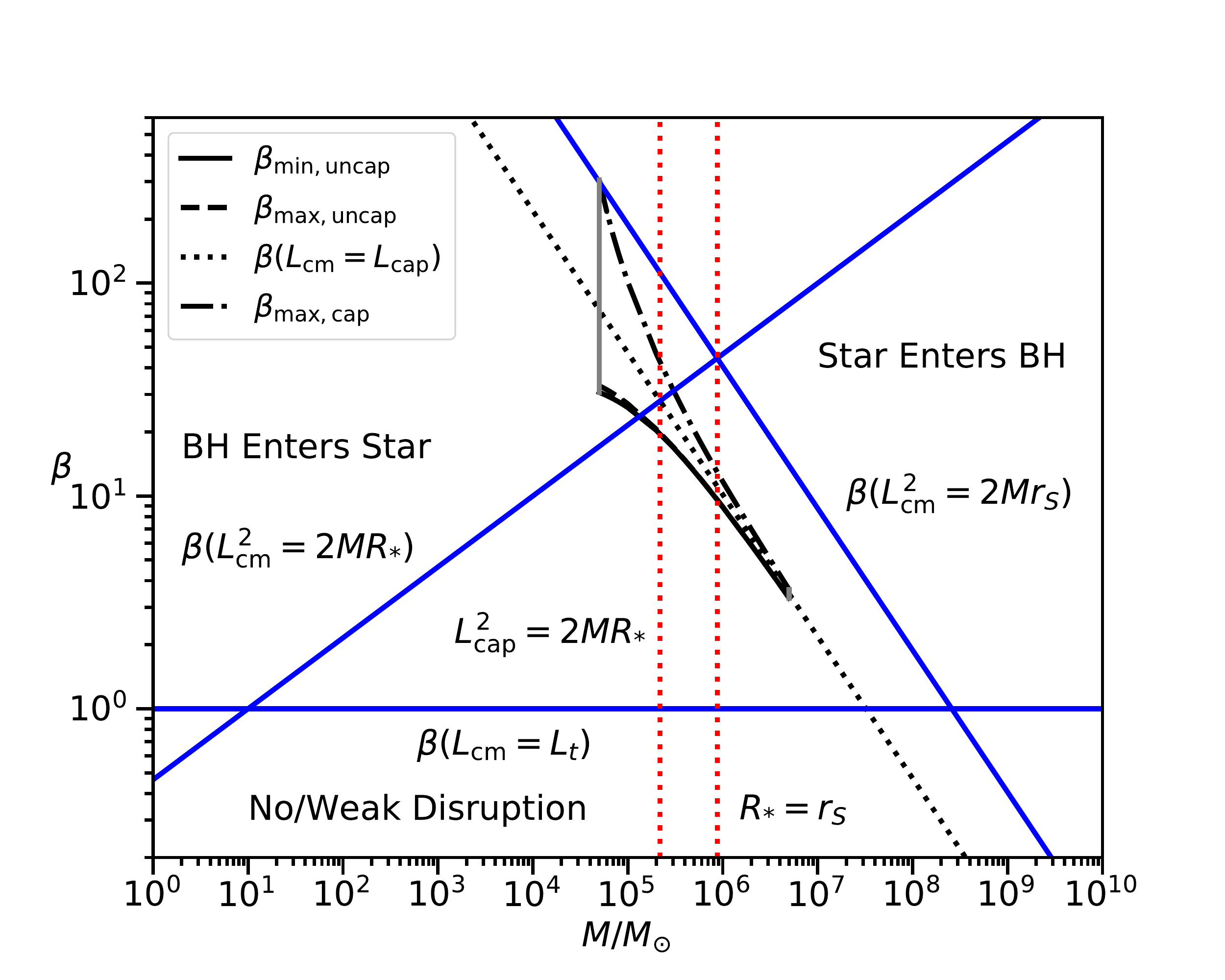}}
\caption{The same as Figure \ref{fig:betamstarovermsol1}, but for stars with $M_* / M_\odot = 10$ and a mass-radius relation of $R_* = R_\odot (M_* / M_\odot)^\alpha$ where $\alpha = 0.57$, which holds for main sequence stars with masses $M_* / M_\odot \gtrsim 1$ \citep{torres10}.}
\label{fig:betamstarovermsol10}
\end{figure*}

Figure \ref{fig:betamstarovermsol1} shows the range of $\beta$ over which we expect prompt self-intersections as a function of the SMBH mass, for Solar-type stars. The minimum (maximum) penetration factor is $\beta_\textrm{min} \simeq 50$ ($\beta_\textrm{max} \simeq 560$) for $M/M_\odot = 10^4$, and decreases to $\beta_\textrm{min} \simeq 5.6$ ($\beta_\textrm{max} \simeq 6.2$) for $M/M_\odot = 10^6$. The event horizon sets the length scale for a given black hole mass. The curves for $\beta_\textrm{min}$ and $\beta_\textrm{max}$ thus converge at higher masses, and the condition $R_* \ll r_t$ breaks down at lower masses $M/M_* \lesssim 10^4$. Figure \ref{fig:betamstarovermsol10} shows the same for stars with $M_* / M_\odot = 10$.

The four black curves in Figures \ref{fig:betamstarovermsol1} and \ref{fig:betamstarovermsol10} divide the space into five regimes of interest: 1) below $\beta_\textrm{min,uncap}$ (solid), there are no prompt self-intersections because $L_\textrm{cm}$ is too large, which leads to a large $r_p$ and thus to a relative precession angle $\delta \phi < 2\pi$; 2) between $\beta_\textrm{min,uncap}$ (solid) and $\beta_\textrm{max,uncap}$ (dashed), none of the debris gets captured, and it precesses sufficiently to produce a self-intersection; 3) between $\beta_\textrm{max,uncap}$ (dashed) and $\beta(L_\textrm{cm}=L_\textrm{cap})$ (dotted), some of the debris gets captured but the CM does not, and the remainder necessarily self-intersects; 4) between $\beta(L_\textrm{cm}=L_\textrm{cap})$ (dotted) and $\beta_\textrm{max,cap}$ (dot-dashed), some of the debris gets captured including the CM, and the remainder necessarily self-intersects; and 5) above $\beta_\textrm{max,cap}$ (dot-dashed), all of the debris gets captured.

Not all of the prompt self-intersections will lead to energetic collisions. The observationally significant self-intersections will occur in regimes 2) and 3), where the CM is not captured. The colliding debris has more mass and a higher density here relative to the other regimes, since the stellar density profile is peaked at the center and decays towards $R_*$. It is these energetic collisions that will lead to qualitatively different behavior and new emission signatures. In contrast, the events in regime 4) will produce weaker self-intersections.

The scaling of $\beta_\textrm{min}$ with $M/M_*$ calculated here exhibits the opposite trend to that found in our order-of-magnitude, analytic estimate in Eq. \ref{eq:betaoomestimate}. For $M_* = 1 M_\odot$, the two roughly agree for $M/M_\odot \sim (0.6 - 1) \times 10^5$; for $M_* = 10 M_\odot$, they roughly agree for $M/M_\odot \sim (2 - 4) \times 10^5$. The analytic estimate is inaccurate for SMBH masses above and below this. For masses above this, though, the range of $\beta$ for a self-intersection shrinks rapidly.

For comparison, \citet{evans15} examined deep encounters with $\beta = 10, 15$ of Solar-type stars incident on a SMBH of mass $M = 10^5 M_\odot$. These values of $\beta$ lie below the range $\beta \simeq 22 - 35$ that we find are required for prompt self-intersection. For these shallower encounters, though, the above Authors still observe efficient debris circularization due to apsidal precession and the rapid formation of an accretion disk.

\section{Hydrodynamic Simulations}
\label{sec:simulations}

In this section, we perform hydrodynamic simulations of deep TDEs of Solar-type stars ($M_* = 1 M_\odot$, $R_* = 1 R_\odot$) by SMBHs of masses $M/M_\odot = (0.6, 1, 1.4) \times 10^5$ (selected from Figure \ref{fig:betamstarovermsol1}) to validate the range of $\beta$ that will produce prompt self-intersections. We use the smoothed-particle hydrodynamics (SPH) code \textsc{phantom} \citep{price18}, which has been successfully used to study a range of TDE phenomena \citep{coughlin15,coughlin16b,bonnerot16,golightly19}. In contrast to these previous studies, we primarily seek to model the orbit and tidal deformation of the debris, and do not seek to accurately model the strong compression, shock, and collision physics. Unlike the geodesic model, though, the simulations incorporate pressure and self-gravity, and the star can experience tidal distortions prior to reaching the tidal radius.

We model the gravity of the SMBH using the generalized Newtonian potential developed by \citet{tejeda13}, which accurately reproduces several features of the Schwarzschild metric. In particular, it captures the orbital frequencies with an error $\lesssim 6 \%$ and exactly reproduces the apsidal precession angle for zero-energy orbits, making it well-suited for studying TDEs. \citet{bonnerot16} successfully used this potential in \textsc{phantom} to examine debris circularization and accretion disk formation. We set the accretion radius of the SMBH to be $1 \%$ larger than the Schwarzschild radius; the SPH particles that cross this radius are captured and removed from the simulation.

We model the star as a polytrope with $\gamma = 5/3$ \citep{stellar04}. To generate the polytrope, we follow the procedure outlined in \citet{coughlin15}, and place the SPH particles on a tightly-packed sphere, stretch the sphere towards a polytropic distribution, and relax the configuration for 10 sound-crossing times to produce a static initial state, which closely matches the known analytical solution.

To simulate an encounter with a penetration factor $\beta$, we place the relaxed star on a Newtonian parabolic orbit at an initial distance $r_i = 3 r_t$ and with pericenter $r_p = r_t / \beta$ (we also ran a few encounters with $r_i = 5 r_t$ and found only small differences). This setup is equivalent to initializing the Newtonian orbit of the stellar CM with the same angular momentum as the corresponding GR orbit; the generalized Newtonian potential should then recover the correct GR orbit. We note, though, that there are several ways to identify a GR orbit with a Newtonian orbit \citep{servin17}.

We use $\sim 10^6$ SPH particles for each simulation. We do not account for shock heating, which is essential for capturing the evolution of the debris after self-intersection. However, our goal here is to test for the existence of the self-intersection, prior to which shock heating should not be important in the midplane. The self-gravity of the SPH particles is implemented using a tree algorithm; we adopt an opening angle criterion of $\theta = 0.5$ to adequately capture the short-range forces \citep{gafton11}. 
%We use a Courant number of $0.3$ for the time steps.
We produce images of the encounters using the visualization tool \textsc{splash} \citep{price07}.

\begin{figure*}
\subfloat[$M/M_\odot = 0.6 \times 10^5$, $\beta = 26$]{\includegraphics[scale=0.2]{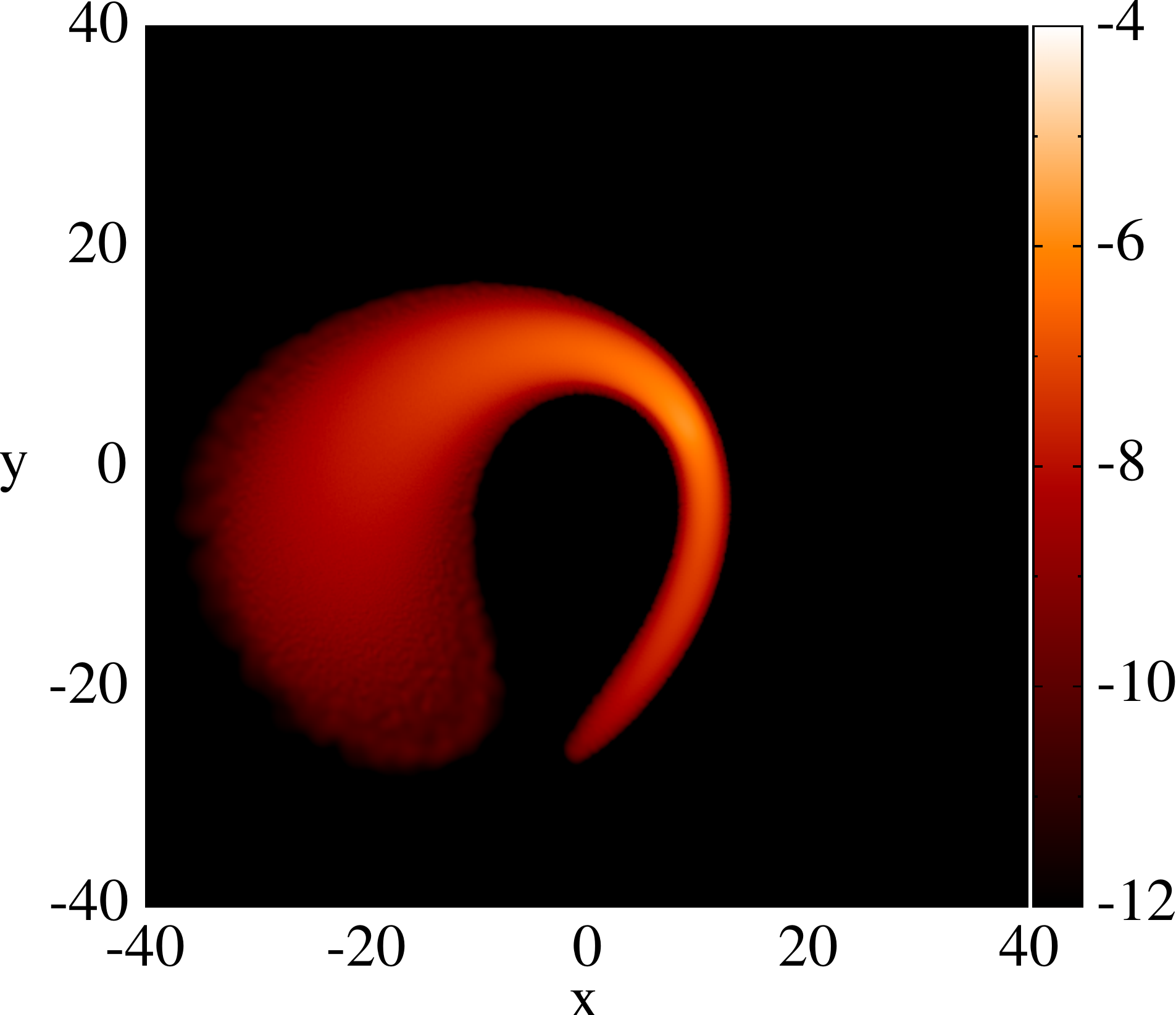}}\hfill
\subfloat[$M/M_\odot = 0.6 \times 10^5$, $\beta = 32$]{\includegraphics[scale=0.2]{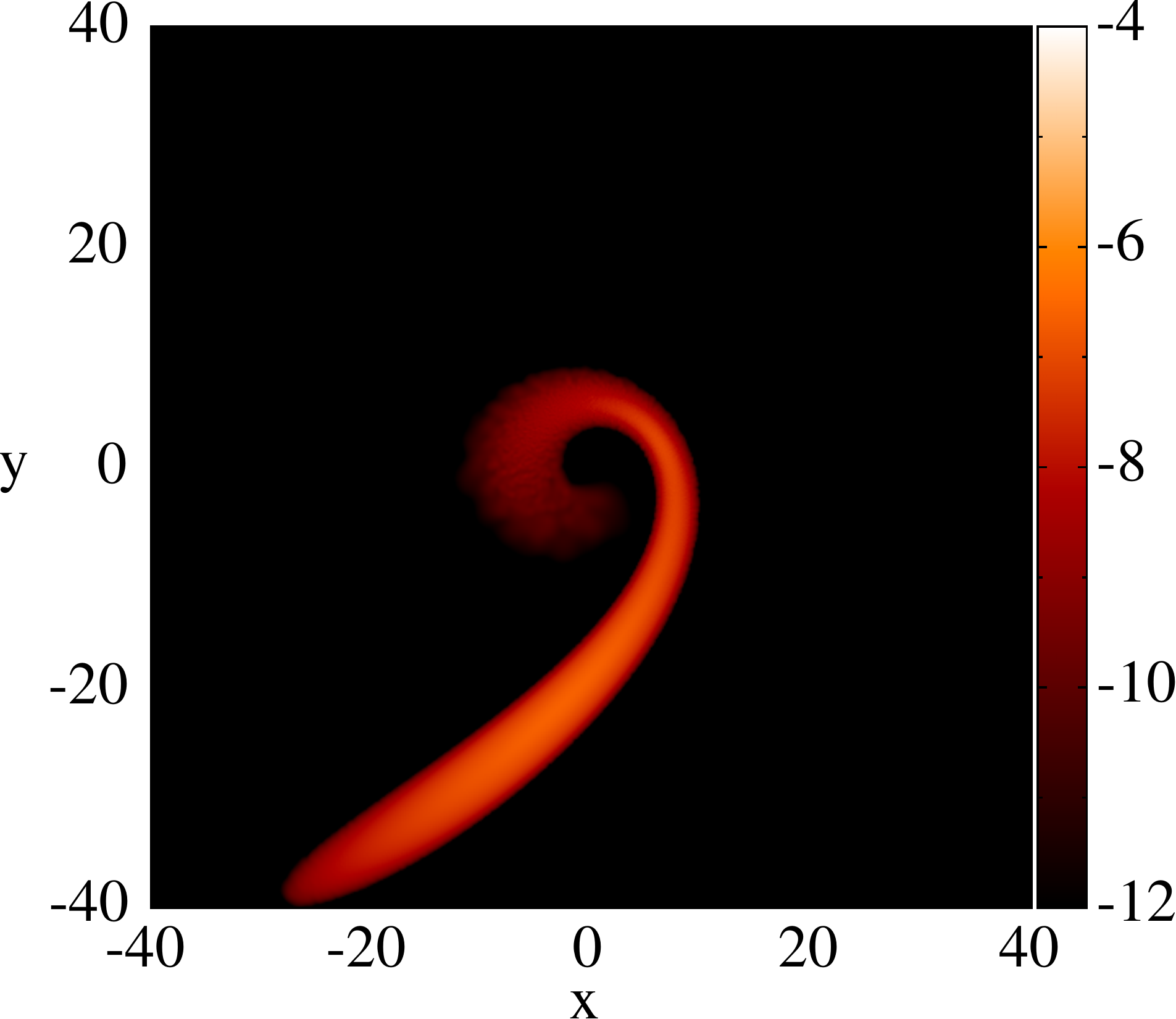}}\hfill
\subfloat[$M/M_\odot = 0.6 \times 10^5$, $\beta = 38$]{\includegraphics[scale=0.2]{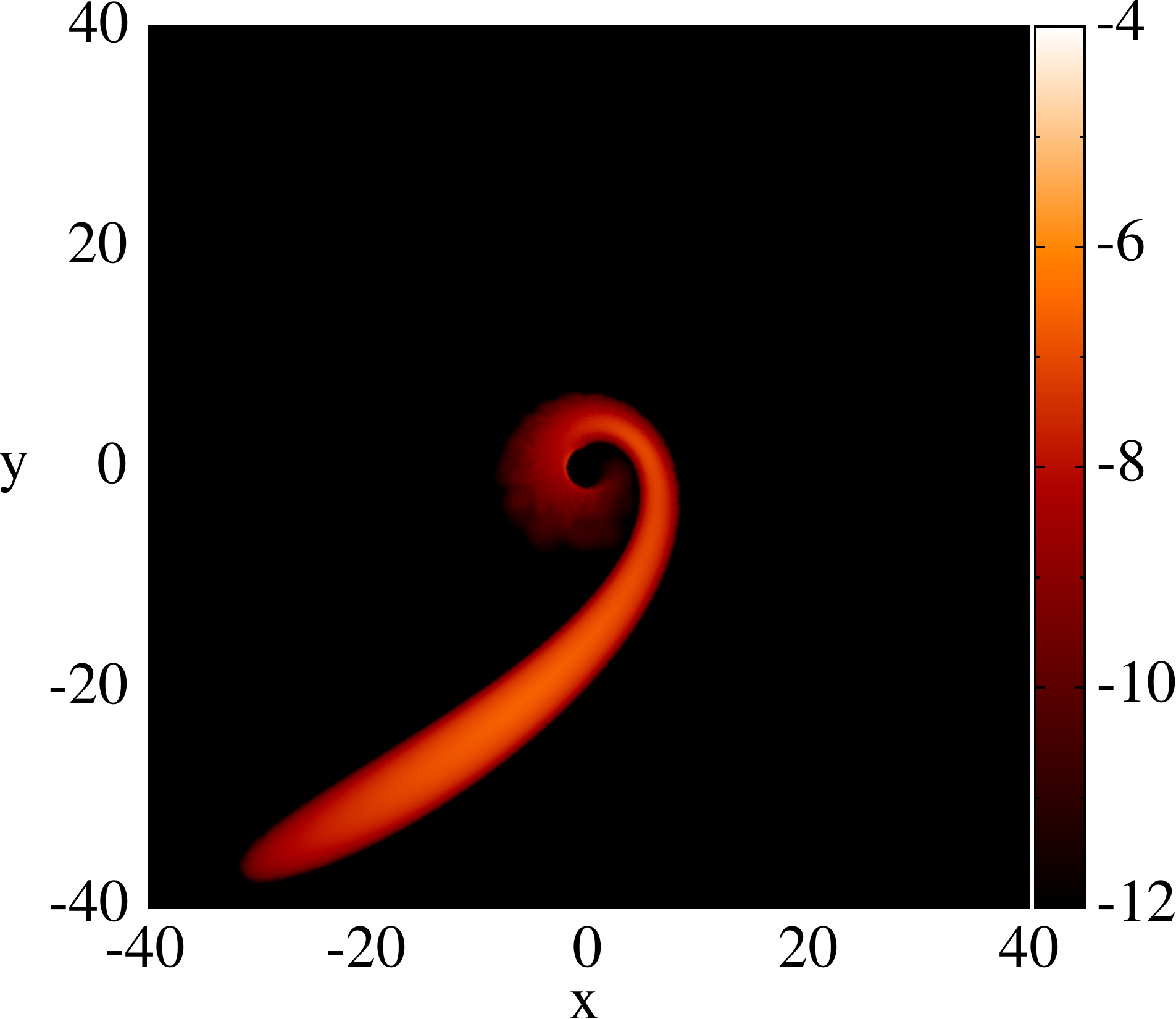}}\hfill
\subfloat[$M/M_\odot = 0.6 \times 10^5$, $\beta = 44$]{\includegraphics[scale=0.2]{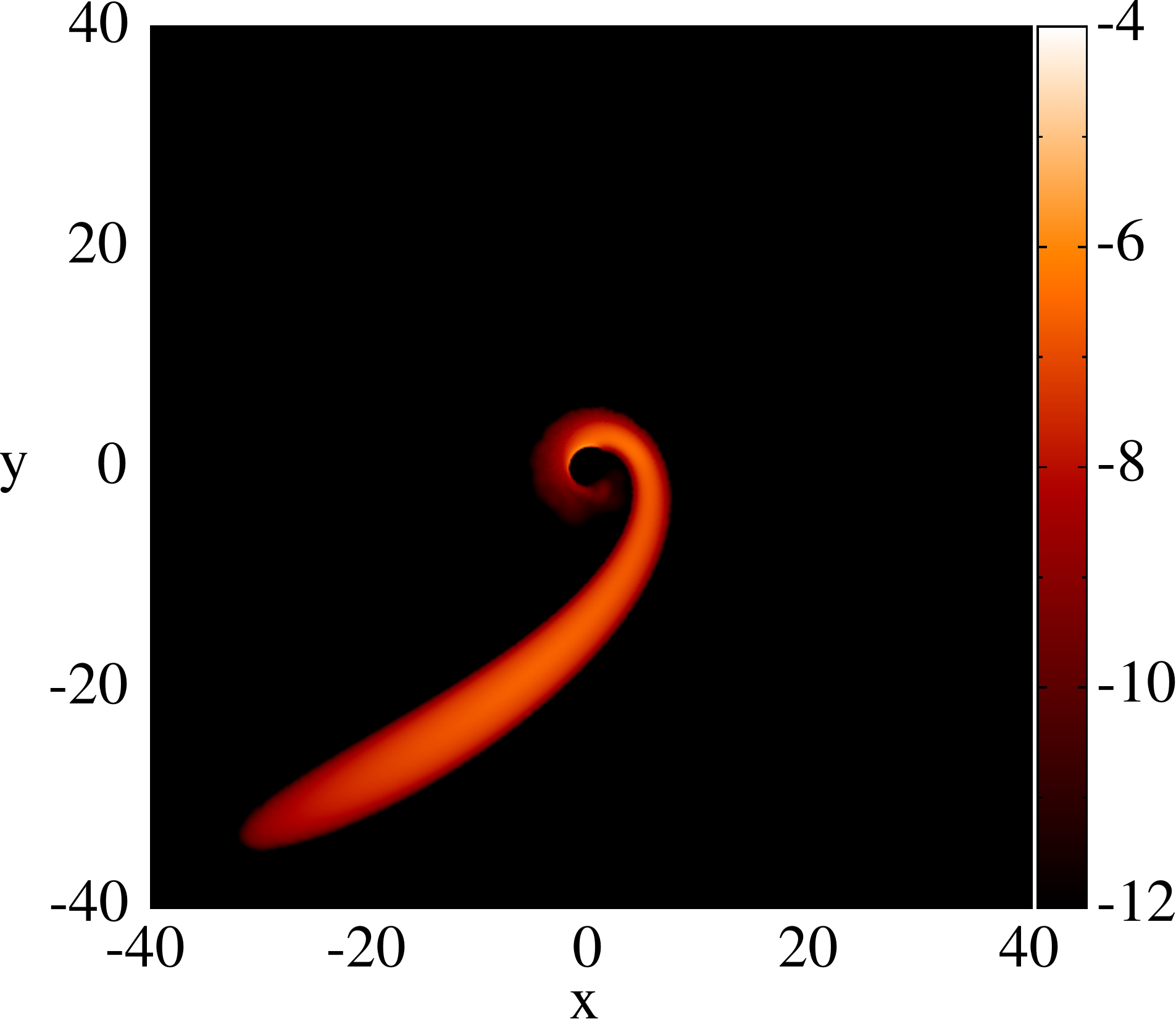}}\\
\subfloat[$M/M_\odot = 10^5$, $\beta = 20$]{\includegraphics[scale=0.2]{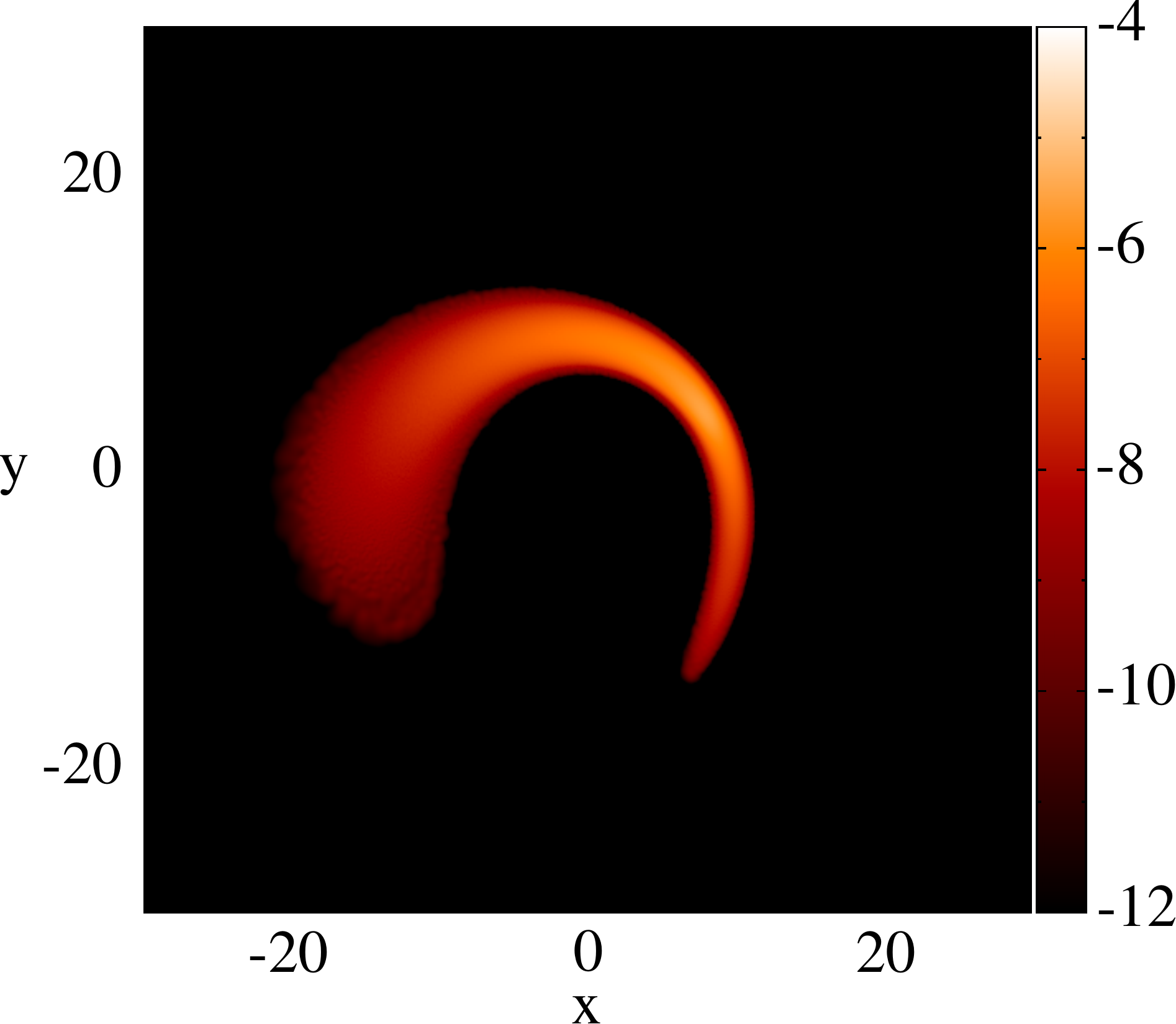}}\hfill
\subfloat[$M/M_\odot = 10^5$, $\beta = 24$]{\includegraphics[scale=0.2]{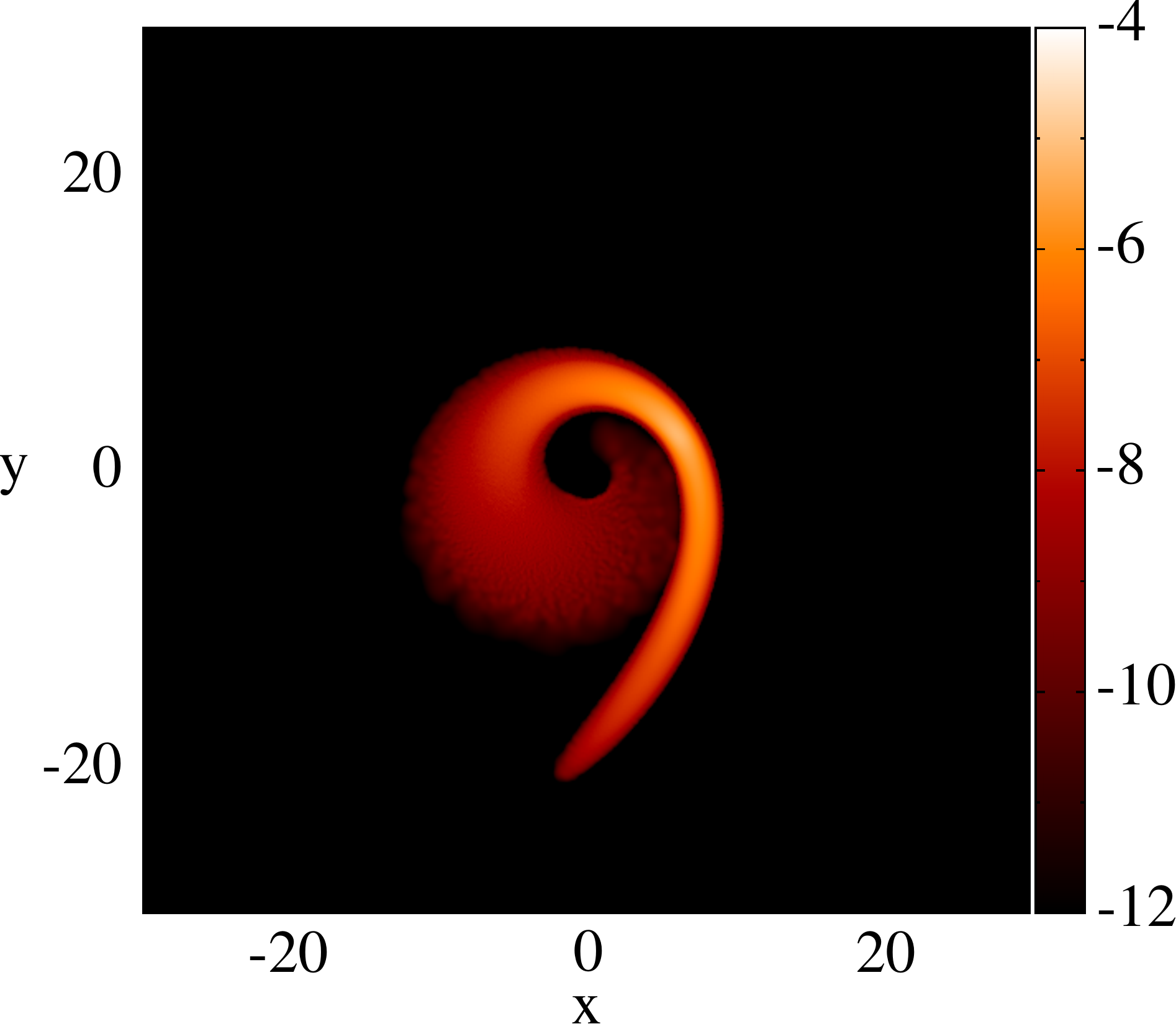}}\hfill
\subfloat[$M/M_\odot = 10^5$, $\beta = 28$]{\includegraphics[scale=0.2]{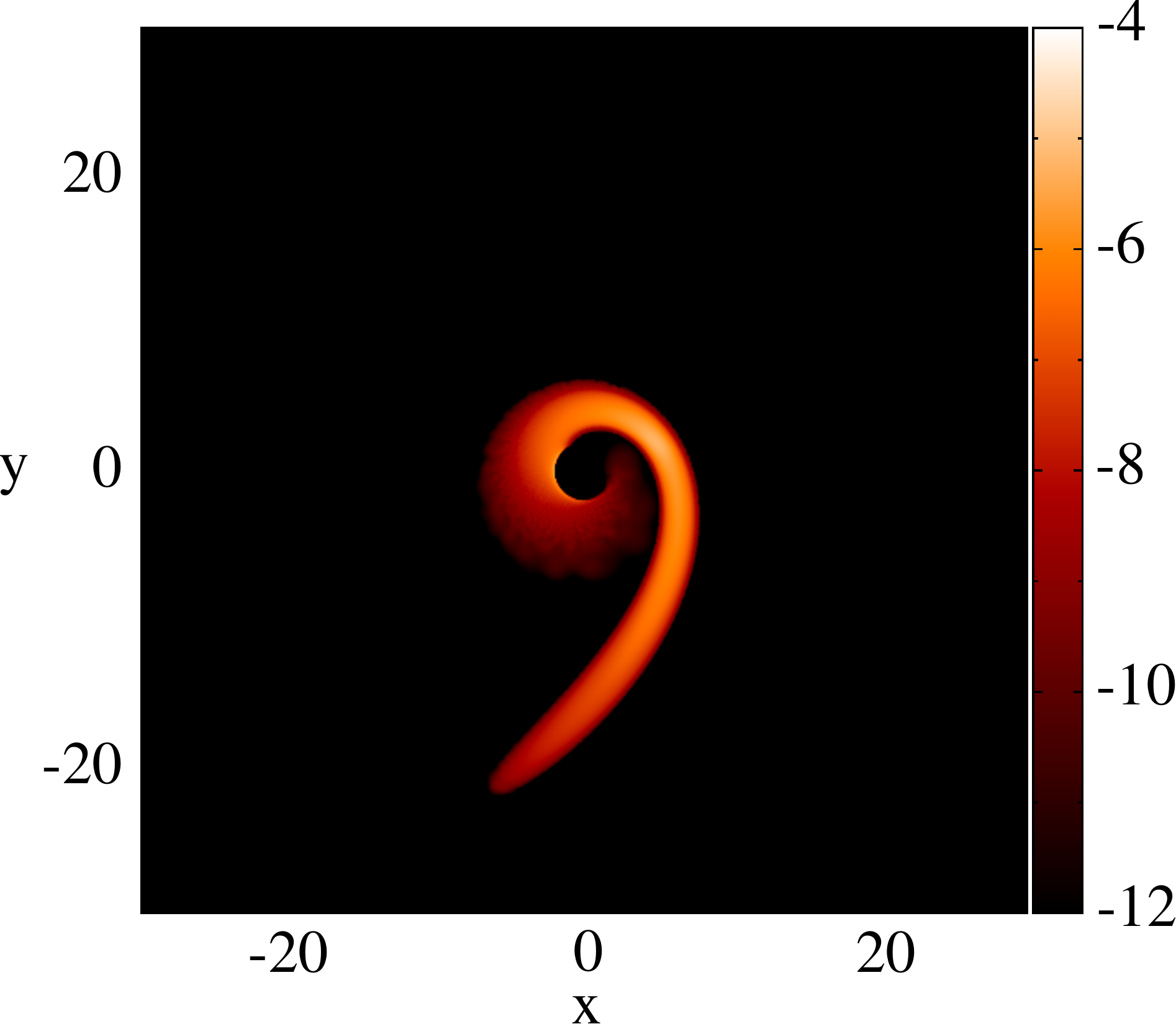}}\hfill
\subfloat[$M/M_\odot = 10^5$, $\beta = 32$]{\includegraphics[scale=0.2]{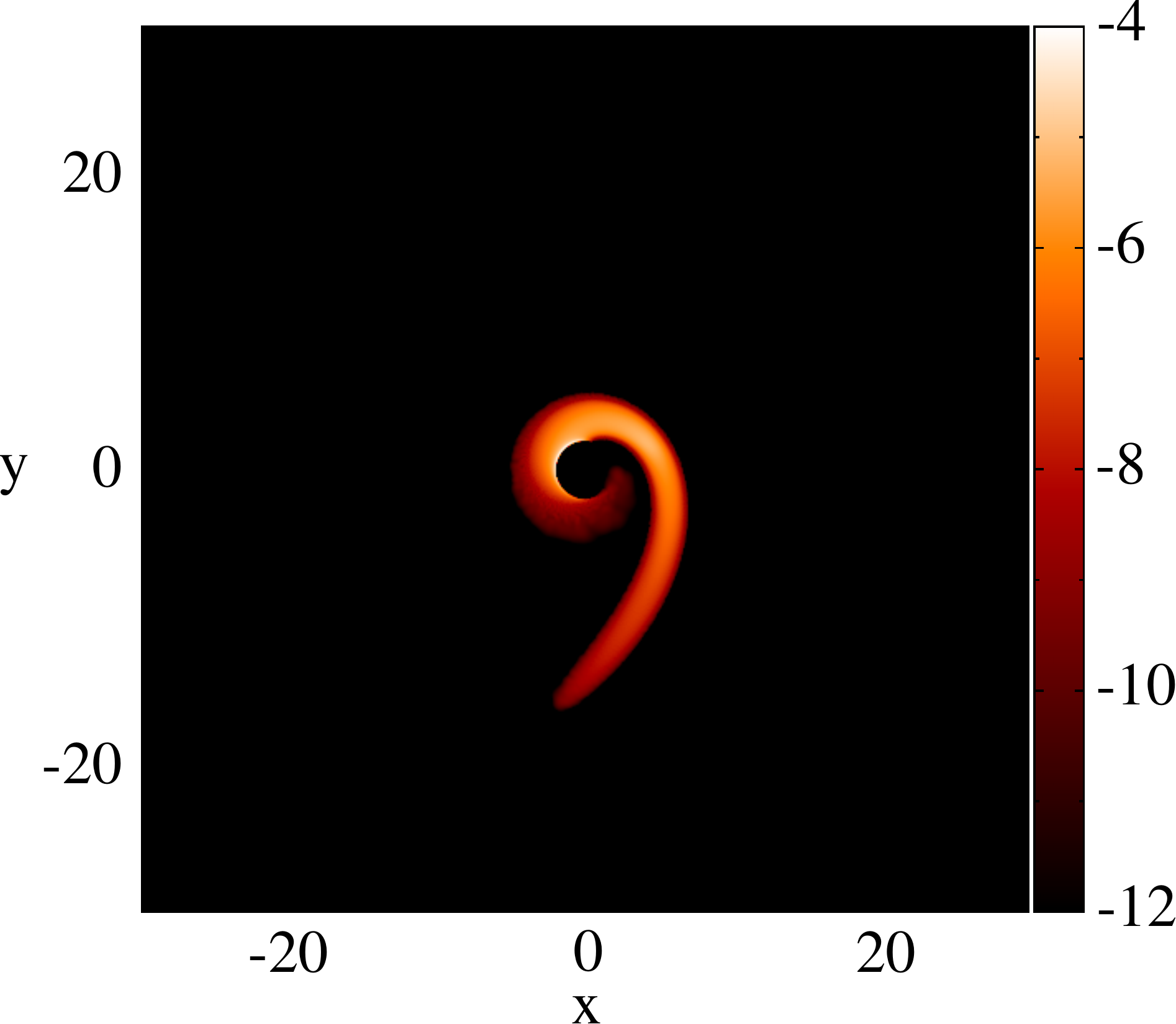}}\\
\subfloat[$M/M_\odot = 1.4 \times 10^5$, $\beta = 18$]{\includegraphics[scale=0.2]{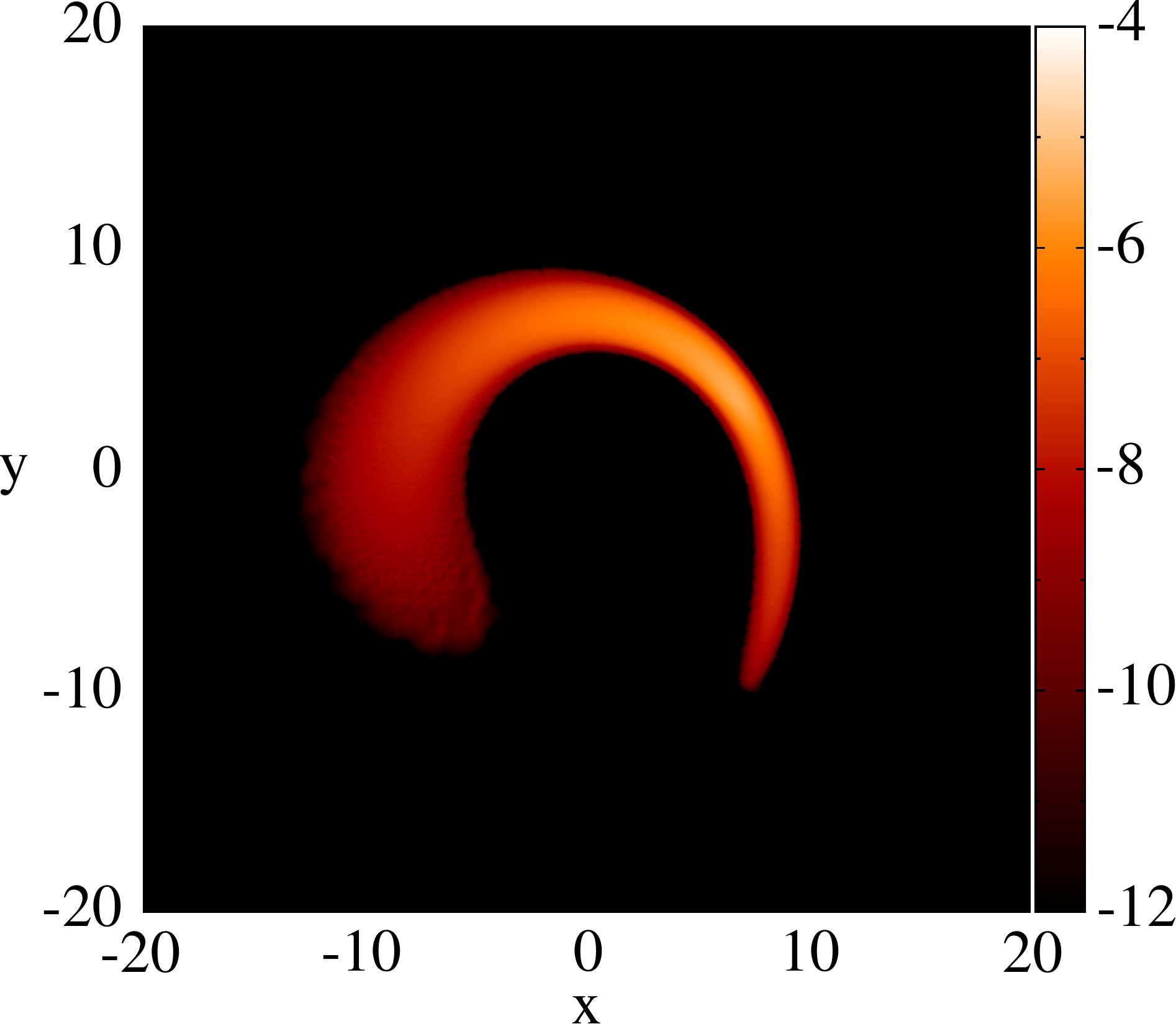}}\hfill
\subfloat[$M/M_\odot = 1.4 \times 10^5$, $\beta = 20$]{\includegraphics[scale=0.2]{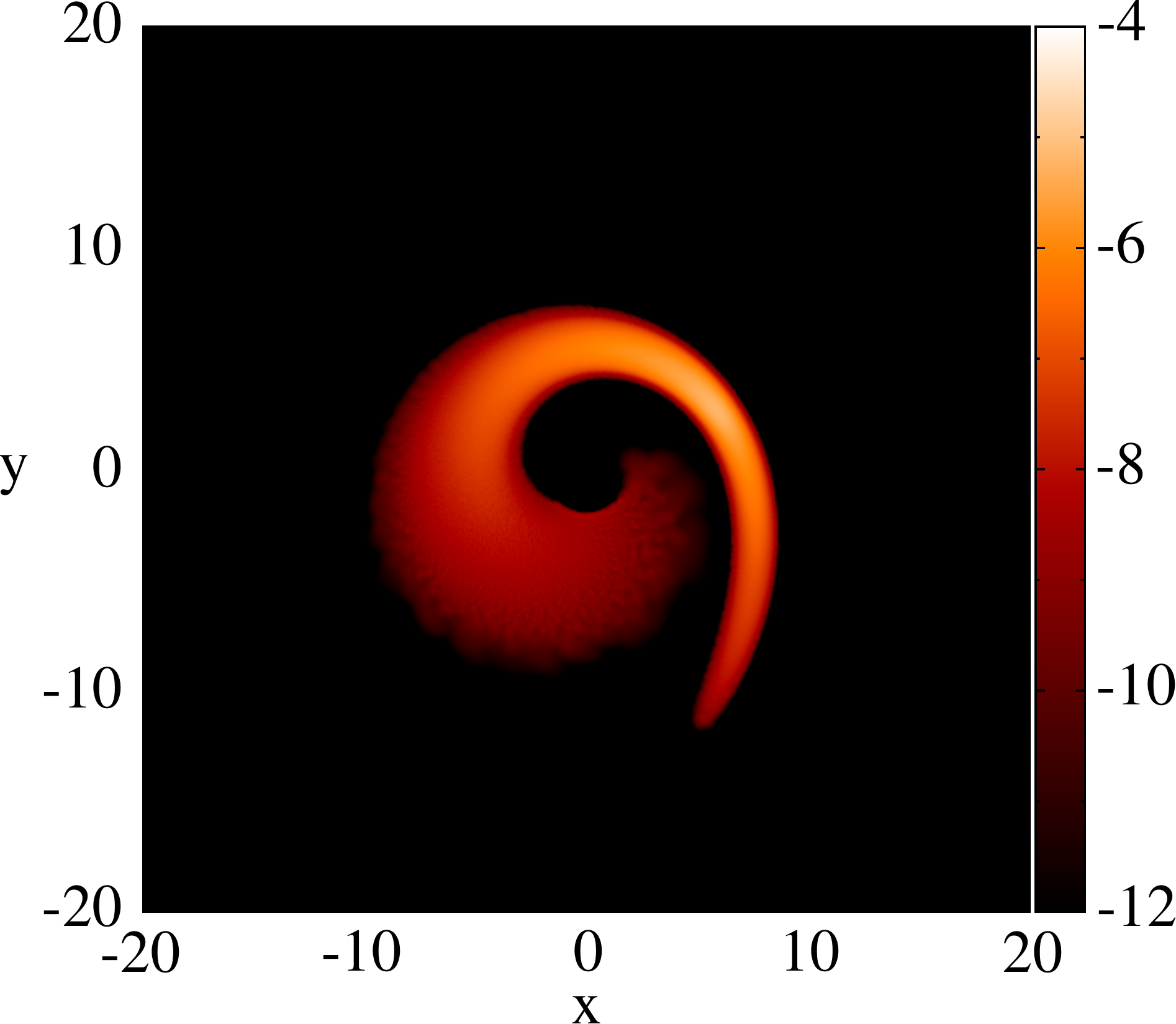}}\hfill
\subfloat[$M/M_\odot = 1.4 \times 10^5$, $\beta = 22$]{\includegraphics[scale=0.2]{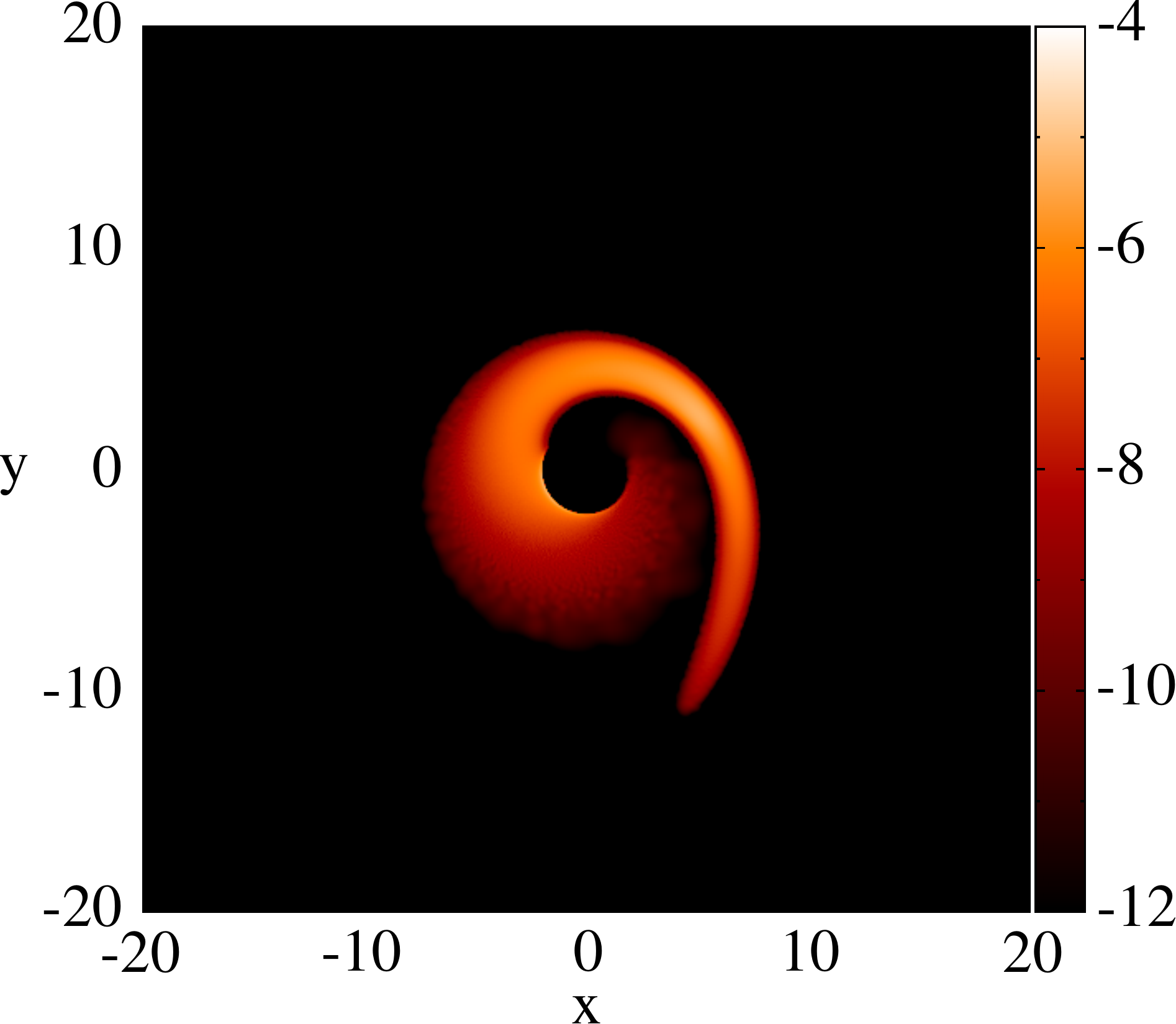}}\hfill
\subfloat[$M/M_* = 1.4 \times 10^5$, $\beta = 24$]{\includegraphics[scale=0.2]{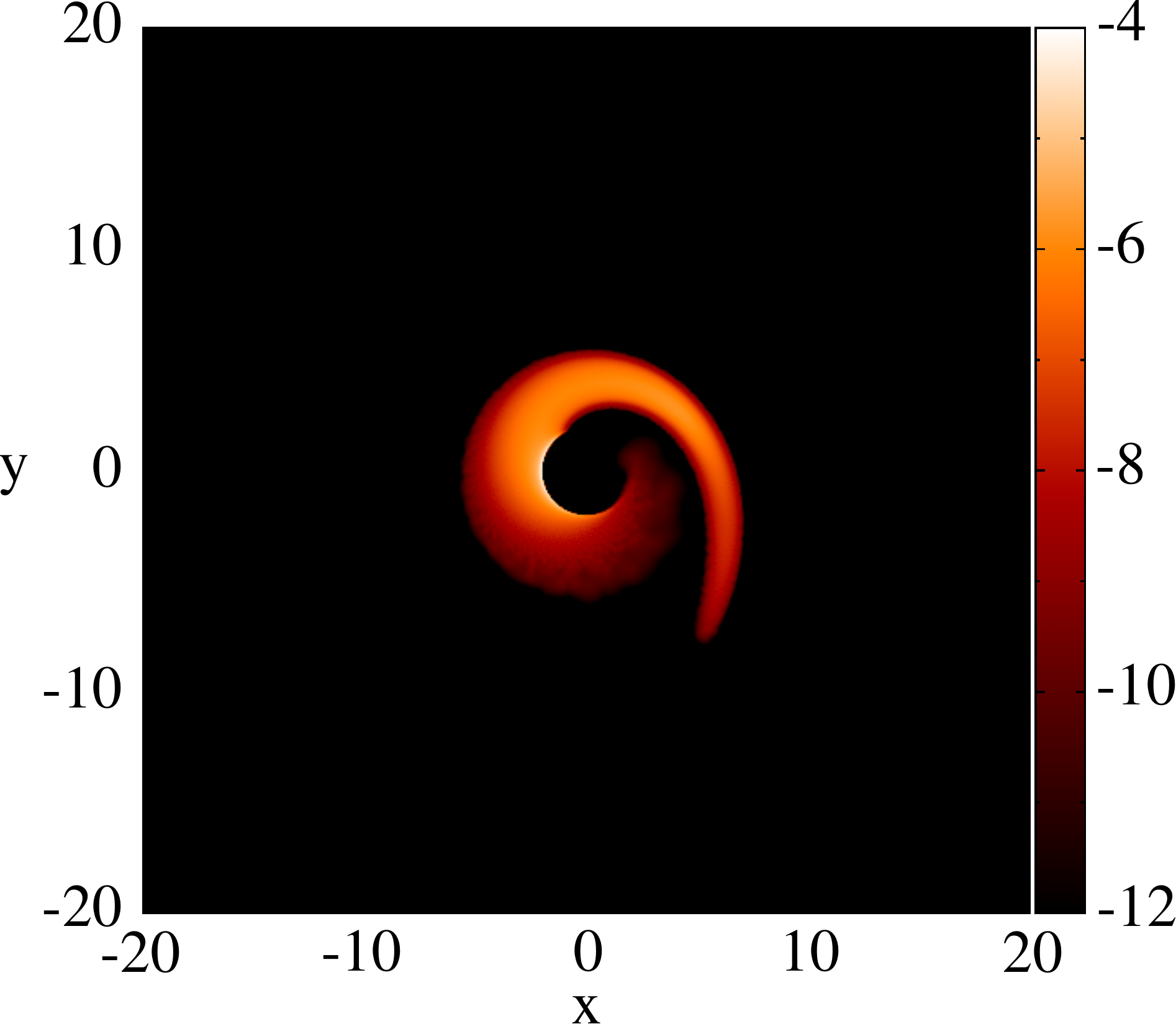}}
\caption{The column density in the $xy$-plane of the disrupted debris near pericenter for Solar-type stars and different values of $M/M_\odot$ and $\beta$. The $xy$ axes are in units of $r_g = GM/c^2$, and the color bar gives the column density on a logarithmic scale. The top row shows $M/M_\odot = 0.6 \times 10^5$, the middle row shows $M/M_\odot = 10^5$, and the bottom row shows $M/M_\odot = 1.4 \times 10^5$. From left to right, the columns show: a,e,i) the debris does not self-intersect; b,f,j) the debris self-intersects; c,g,k) the CM of the star is captured and the remaining debris weakly self-intersects; d,h,l) most of the debris is captured.}
\label{fig:sims}
\end{figure*}

Figure \ref{fig:sims} shows the column density in the $xy$-plane of the disrupted debris near pericenter for Solar-type stars and different values of $M/M_\odot$ and $\beta$. For $M/M_\odot = (0.6, 1, 1.4) \times 10^5$, the SPH simulations give $\beta_\textrm{min} \simeq 27, 22, 19$ for self-intersection. These values are in close agreement with those from the geodesic model ($\beta_\textrm{min} \simeq 28, 22, 19$; Figure \ref{fig:betamstarovermsol1}), with a slightly shallower scaling of $\beta_\textrm{min}$ with $M/M_\odot$. The SMBHs in the simulations begin capturing the stellar CM around $\beta = 38, 28, 22$ and capture most of the debris by $\beta = 44, 32, 24$, which are again in close agreement with the values of $\beta(L_\textrm{cm} = L_\textrm{cap})$ and the region between $\beta(L_\textrm{cm} = L_\textrm{cap})$ and $\beta_\textrm{max}$ predicted by the geodesic model. The simulations also confirm that the higher mass SMBHs require a smaller range of $\beta$ for self-intersection.

These simulations confirm the basic predictions of the geodesic model of Section \ref{subsec:geodesic} and the more crude, analytic estimates of Section \ref{subsec:analytic}. The initial tidal distortion of the star before it reaches the tidal radius could explain the slightly smaller values of $\beta$ for low SMBH masses for which the SPH simulations yield self-intersections; the star is slightly elongated at the time it reaches $r_t$, so it does not need to be subsequently stretched by the amount predicted by the ``frozen-in'' approximation in order to self-intersect. We note, however, that the minimum penetration factor for self-intersection that we find here ($\beta_\textrm{min} \simeq 22$ for $M = 10^5 M_\odot$) is still well above the value of $\beta = 10$ found numerically by \citet{evans15}. The origin of this discrepancy is unclear, though it may be due to the softer polytropic equation of state that those authors employed.

\section{Discussion}
\label{sec:discussion}

In this paper, we examined a new regime of ultra-deep TDEs of main sequence stars in which the disrupted debris promptly intersects itself at the first pericenter passage. This is in contrast to canonical TDEs in which the debris gradually intersects itself following the return of the debris to pericenter \citep{rees88}. We calculated the range of SMBH masses $M$ and penetration factors $\beta$ for which these prompt self-intersections occur, using 1) a rough, order-of-magnitude, analytic estimate (Section \ref{subsec:analytic}); 2) a geodesic model under the impulse and ``frozen-in'' approximations (Section \ref{subsec:geodesic}); and 3) SPH simulations (Section \ref{sec:simulations}). In all three of these approaches, we demonstrated that one needs a combination of large penetration factor ($\beta \gtrsim 10$) and small black hole mass ($M \lesssim 10^6 M_{\odot}$) in order for the debris to experience extreme tidal distortion as it plunges within the tidal sphere, to travel through large precession angles at pericenter ($\gtrsim 2\pi$), and to avoid plunging into the SMBH.

These ultra-deep TDEs will produce observable electromagnetic (EM) and gravitational wave (GW) signals closely spaced in time. We provide rough estimates for the multimessenger signatures of such an event here, but leave a detailed analysis to a future study. We then briefly discuss the detection prospects for these events.

We begin with an estimate of the prompt radiation emitted in the self-intersection. We model the debris stream as a cylinder of length $l \sim r_p$ and radius $r \sim R_*$. A smaller portion $l_c < l$ of the debris stream is shocked in the intersection. This is a strong shock, which deposits a thermal energy density of order $\rho v_s^2$, where $\rho \sim 3M_*/4\pi R_*^3$ is the debris mass density (assumed constant) and $v_s$ is the shock velocity, which roughly equals the debris velocity at pericenter, $v_s \sim v_p \sim (GM/r_p)^{1/2}$. The stream is optically thick, so only thermal energy in the surface layers can diffuse out before the remainder is degraded by adiabatic expansion. The optical depth of the surface layer is $\tau = \rho \kappa \Delta R$, where $\Delta R$ is its thickness and $\kappa \sim 0.4 \textrm{ cm}^2 \textrm{g}^{-1}$ is the opacity, assumed to be dominated by electron scattering. If we impose the condition that the photon diffusion time through the layer, $t_\textrm{diff} \sim \tau \Delta R/c$, must be shorter than the debris expansion time, $t_\textrm{exp} \sim R_*/v_p$, then we obtain $\Delta R \lesssim ( c R_* / \rho \kappa v_p )^{1/2}$. The thermal energy in the surface layer is then $\Delta E \sim (2 \pi R_* l_c \Delta R) (\rho v_s^2) \sim  2 \pi l_c ( R_*^3 v_p^3 c \rho / \kappa )^{1/2}$. For $M \sim 10^5 M_\odot$, solar-type stars, $r_p \sim 5r_g$, and an interaction length of $l_c \sim R_*$, this is $\Delta E \sim 1.3 \times 10^{48}$ erg. The light crossing time, $t_\textrm{lc} \sim R_*/c \sim 2.3$ s, is much larger than the photon diffusion time, $t_\textrm{diff} \sim 1.2 \times 10^{-5}$ s; the former thus determines the timescale over which the energy will appear to be radiated to a distant observer, which gives an observed luminosity of order $L \sim 5.5 \times 10^{47}$ erg/s. The effective temperature $T$ can also be estimated by $L = (2 \pi R_* l_c) \sigma T^4$, giving $T \sim 2.4 \times 10^7$ K ($kT \sim 2.1$ keV). The collision will thus produce a short, bright flare in X-ray wavelengths.

As discussed above, the strong shock from self-intersection converts much of the kinetic energy $(\sim \rho v_p^2)$ into thermal energy, which is comparable to the gravitational binding energy. Following this, the fate of the debris is uncertain; some will plunge into the black hole, some will form a disk and accrete onto the black hole through viscous processes, and some will be ejected. The debris that plunges into the black hole will not emit an observable EM signature. The debris that accretes will have a prompt accretion phase over the initial viscous timescale, $t_\textrm{visc}$, and a delayed accretion phase over a longer timescale. For the prompt phase, we can roughly estimate the viscous time $t_\textrm{visc}$ using the $\alpha$-viscosity prescription for thin disks developed by \citet{shakura76}, which gives $t_\textrm{visc} \sim \alpha^{-1} (h/r)^{-2} P$ where the debris has scale height $h$, radial extent $r$, orbital period $P$, and dimensionless viscous parameter $\alpha$, though we note that the accreting debris is well outside the thin disk regime. For $h \sim R_*$, $r \sim r_p$, $P \sim 2\pi (r_p^3 / GM)^{1/2}$, $\alpha \sim 0.1$, and the stellar and black hole parameters above, we find $t_\textrm{visc} \sim 6.5$ mins. The accretion rate is then $\dot{M} \sim \eta M_* / t_\textrm{visc}$, where $\eta$ is the fraction of debris that is promptly accreted. For $\eta \sim 0.1 - 0.5$, this yields $\dot{M} \sim (0.8 - 4) \times 10^4 M_\odot/$yr. \citet{evans15} found similar accretion rates and timescales for the prompt accretion episodes in their simulations, though for shallower encounters than considered here. We can estimate the accretion luminosity as $L \sim \epsilon \dot{M} c^2$ ergs/s, where $\epsilon$ is the radiative efficiency. For even a modest $\epsilon \sim 0.1$, the emission is highly super-Eddington, $L/L_\textrm{Edd} \sim 10^7 - 10^8$. The radiation may drive an outflow from the disk, yielding an observed luminosity that is Eddington-limited, or it may be highly beamed in the form of a jet that does not unbind the disk, yielding an intrinsic accretion rate close to $\dot{M}$ \citep{coughlin14}. For the delayed phase, it is unclear whether the debris accretes with the characteristic $t^{-5/3}$ decay of conventional TDEs \citep{kawana18,anninos18}, or at a roughly constant rate \citep{evans15}, or exhibits an altogether different behavior. The debris that is ejected, both from the self-intersection and the super-Eddington accretion, can produce an afterglow when it collides with matter surrounding the galactic nucleus.

There is another potential EM signature that may occur in a deep encounter even before the initial pericenter passage. A star approaching a SMBH in a deep encounter may experience strong tidal compression as it approaches pericenter, which will generate a shock wave that propagates to the surface, heats the outer layers, and produces an X-ray signature \citep{kobayashi04,guillochon09,yalinewich19}. If compression does occur, it may ignite nuclear reactions \citep{carter82,bicknell83}. Early studies of the X-ray breakout examined encounters with black hole masses $M \sim 10^6 M_\odot$, solar-type stars, and $\beta \sim 5 - 10$, and found luminosities $L \sim 10^{42} - 10^{44}$ ergs/s at average photon energies $E \sim 1 - 4$ keV \citep{kobayashi04,guillochon09}. Recently, \citet{yalinewich19} lowered this prediction to $L \sim 10^{41}$ ergs/s at $E \sim 1 - 10$ keV, the reduction arising from a more rapid drop in the shock velocity with increasing distance (decreasing density) from the midplane. For ultra-deep encounters, the debris may even spread sufficiently before pericenter to prevent X-ray breakout altogether \citep{evans15}.

It is interesting to compare the two types of X-ray flares discussed above: one from self-intersection in ultra-deep encounters, and one from tidal compression in general deep encounters. The self-intersection flare is $\sim 3$ orders of magnitude brighter than the compression flare. The higher brightness arises because the specific kinetic energy budget available for shock heating is much larger for self-intersection, $v_p^2 \sim GM/r_p$, than tidal compression, $v^2 \simeq \beta^2 (GM_* / R_*)$ \citep{carter82}. Even if the self-intersection shock degrades by $\sim 2 - 3$ orders of magnitude like the compression shock, it will still produce an X-ray flare at least as bright as the most optimistic estimates from the compression shock.

Deep encounters will also produce two types of GW signatures. The first signature arises from the orbital motion of the debris near pericenter \citep{kobayashi04,guillochon09}, including when the debris is fully captured \citep{east14}. The GW frequency and strain for these events are roughly \citep{kobayashi04,guillochon09}
\begin{align}
f &\sim \left(\frac{GM}{r_p^3}\right)^{1/2} \sim (6.3 \times 10^{-4} \textrm{ Hz}) \beta^{3/2} m_*^{1/2} r_*^{-3/2} \\
h &\sim \frac{GM_* r_S}{c^2 d r_p} \sim 4.4 \times 10^{-23} \beta d_{10}^{-1} m_*^{4/3} r_*^{-1} M_5^{2/3}
\end{align}
where $d_{10} \equiv d / (10 \textrm{ Mpc})$ with $d$ being the distance to the event. For the distance $d_{10} = 1$ and the parameters in Figure \ref{fig:betamstarovermsol1} (i.e. the stellar parameters, the range of SMBH masses, and the values of $\beta_\textrm{min}$), we find $f \sim (2.2 - 0.06) \times 10^{-1}$ Hz and $h \sim (0.47 - 1.1) \times 10^{-21}$. These lie outside the sensitivity of both advanced LIGO \citep{ligo15} and LISA \citep{lisa2017}, but are accessible to currently proposed GW detectors \citep{moore15}, notably those in the decihertz range \citep{decigo06,decigo17}. For the parameters in Figure \ref{fig:betamstarovermsol10}, we find $f \sim (4.7 - 0.17) \times 10^{-2}$ Hz and $h \sim (0.49 - 1.1) \times 10^{-20}$. These fall marginally within the sensitivity of LISA. The second GW signature arises from the tidal deformation of the star itself \citep{guillochon09,stone13}. For our parameter range, the frequency and strain are roughly \citep{stone13}
% \citet{stone13} find that three-dimensional desynchronization is important for encounters with $\beta \gtrsim \beta_d \equiv 1.6 (M/M_*)^{1/12}$, a range in which our ultra-deep TDEs fall. In this case, the frequency and the dominant contribution to the strain in the absence of reflection symmetry are roughly \citep{stone13}
\begin{align}
f &\sim (3.8 \times 10^{-5} \textrm{ Hz}) \beta^4 m_*^{1/2} r_*^{-3/2} \\
h_+ &\sim 1.2 \times 10^{-24} \beta^{-2} d_{10}^{-1} m_*^{11/6} r_*^{-1} M_5^{1/3}
\end{align}
% For the distance $d_{10} = 1$ and the parameters in Figure \ref{fig:betamstarovermsol1}, we find $f \sim (2.3 - 0.00038) \times 10^{2}$ Hz and $h \sim (0.010 - 3.8) \times 10^{-24}$. For the parameters in Figure \ref{fig:betamstarovermsol10}, we find $f \sim (1.5 - 0.00020) \times 10^{1}$ Hz and $h \sim (0.0086 - 3.4) \times 10^{-22}$. 
These signals are not detectable by advanced LIGO or LISA, and fall only marginally within the sensitivity of future GW detectors.

Ultra-deep TDEs are statistically rare due to the large penetration factors required to achieve them. The probability of a prompt self-intersection depends on the BH mass function over our mass range and the state of the stellar loss cone at a given BH mass, namely the degree to which it is ``full'' (``pinhole'' regime) or ``empty''  (``diffusive'' regime) \citep{frank76,lightman77}. We assume a uniform distribution for the BH mass function over our mass range, though we note that the scaling depends on the model assumptions used to construct it \citep{stone16,kochanek16,vanvelzen18}. The TDE rate is then $\dot{N}_\textrm{TDE} \sim 10^{-5}$ Mpc$^{-3}$ yr$^{-1}$ for SMBHs with masses $M \sim (10^4 - 10^6) M_\odot$ \citep{stone16}. We also assume that the loss cone is ``full'' over our mass range, which likely holds for SMBHs with masses $M \sim (10^5 - 10^6) M_\odot$ \citep{stone16}. The incident stars will then have penetration factors distributed according to the PDF $f_B(\beta) = \beta^{-2}$ \citep{luminet90,stone16,kochanek16}. The probability of a significant prompt self-intersection over our mass range is then simply $P_{SI} \simeq \beta_\textrm{min}^{-1} - \beta(L_\textrm{cm} = L_\textrm{cap})^{-1} \lesssim 1 \%$, yielding an ultra-deep TDE rate of $\dot{N}_\textrm{UD} \lesssim 10^{-7}$ Mpc$^{-3}$ yr$^{-1}$.

The X-ray flare from self-intersection evolves quickly, akin to a gamma-ray burst (GRB), and distinguishes ultra-deep TDEs from conventional ones. We estimate its detectability. We use the current parameters for a flat $\Lambda$CDM cosmology \citep{hinshaw13}. For simplicity, we round the numbers in our emission estimate above, and consider a monochromatic source with emitted luminosity $L_e \sim 5 \times 10^{47}$ erg/s, energy $E_e \sim 2$ keV (frequency $\nu_e = E_e / h$), and duration $\Delta t_e \sim 2$ s. If the source is at redshift $z$, the detector will observe the event with energy $E_o = E_e / (1+z)$ (frequency $\nu_o = \nu_e / (1+z)$) over a duration $\Delta t_o = (1+z) \Delta t_e$. The flux at the detector is $S_o = L_e / 4\pi D_L^2$, where $D_L$ is the luminosity distance. We take the integration time at the detector to be $\Delta t_\textrm{int} \sim \Delta t_e$. The number of counts measured at the detector is then $n = S_o \Delta t_\textrm{int} / E_o A$,  where $A$ is the effective area of the detector. The background is low since our integration time is short, so a signal of $n = 10$ counts provides a $\sim 3\sigma$ detection. The maximum luminosity distance $D_L$ that can be observed at this level is given by $D_L / (1+z)^{1/2} = ( L_e \Delta t_\textrm{int} A / 4\pi n E_e )^{1/2}$, which yields a corresponding comoving volume $V_C$. The detection rate for ultra-deep TDEs is then $R = \dot{N}_\textrm{UD} V_C \Omega / 4\pi$, where $\Omega$ is the detector field of view.

We estimate the detection rate using parameters characteristic of two X-ray missions with the required energy range: the Swift X-ray Telescope (XRT) \citep{burrows05}, and the upcoming extended Roentgen Survey with an Imaging Telescope Array (eROSITA) \citep{merloni12}. The X-ray flare is below the energy range of Swift BAT \citep{krimm13}. For Swift XRT ($A \sim 125$ cm$^2$, $\Omega = 23.6$ arcmin$^2$), we find a limiting distance $D_L \sim 9$ Gpc with detected energy $E_o \sim 0.9$ keV and a detection rate $R \sim 0.1$ yr$^{-1}$, which makes a serendipitous detection unlikely. For eROSITA ($A \sim 1000$ cm$^2$, $\Omega = 0.833$ deg$^2$), we find $D_L \sim 36$ Gpc with $E_o \sim 0.4$ keV and $R \sim 3$ yr$^{-1}$, which is slightly more promising.

In addition, the potential late-time dynamics may be observable in wide field optical surveys. Though $\dot{N}_\textrm{UD}$ is small, current wide-field surveys such as the Zwicky Transient Facility (ZTF) \citep{bellm19} and upcoming surveys like the Large Synoptic Survey Telescope (LSST) \citep{ivezic19} should collectively detect hundreds to thousands of TDEs, yielding a few to dozens of these rare events.

\section*{Acknowledgements}

We thank the anonymous referee for providing thorough comments. We thank Eliot Quataert, Jeremy Hare, and Alex Krolewski for helpful discussions. This research used resources of the National Energy Research Scientific Computing Center, a Department of Energy Office of Science User Facility supported by the Office of Science of the U.S. Department of Energy under Contract No. DE-AC02-05CH11231. This work was supported by the National Science Foundation under Grant No. 1616754. ERC was supported by the National Aeronautics and Space Administration through the Einstein Fellowship Program, Grant PF6-170170. CJN is supported by the Science and Technology Facilities Council (STFC) (grant number ST/M005917/1).

%%%%%%%%%%%%%%%%%%%%%%%%%%%%%%%%%%%%%%%%%%%%%%%%%%



\bibliographystyle{mnras}
\bibliography{references} % if your bibtex file is called references.bib

%% Alternatively you could enter them by hand, like this:
%% This method is tedious and prone to error if you have lots of references
%\begin{thebibliography}{99}
%\end{thebibliography}

%%%%%%%%%%%%%%%%%%%%%%%%%%%%%%%%%%%%%%%%%%%%%%%%%%

%%%%%%%%%%%%%%%%% APPENDICES %%%%%%%%%%%%%%%%%%%%%

\appendix

%If you want to present additional material which would interrupt the flow of the main paper, it can be placed in an Appendix which appears after the list of references.

\section{Orthonormal Tetrad}
\label{sec:appendixa}

In the paper, we examined a deep TDE in the Schwarzschild metric, and used an orthonormal tetrad (ONT) $\lambda_{(\mu)}$ that is parallel-propagated along a timelike geodesic to define Fermi Normal Coordinates (FNCs). In this appendix, we give the components of the tetrad elements in the Schwarzschild coordinate basis; these were presented in earlier work \citep{luminet85,brassart10}, and obtained from the more general expressions derived by \citet{marck83} in the Kerr metric.

The zeroth element of the tetrad is simply the 4-velocity of the center-of-mass (CM) of the star, $\lambda_{(0)} = u_\textrm{cm} = \dot{t} \partial_t + \dot{r} \partial_r + \dot{\phi} \partial_\phi$. We considered the CM geodesic to move in the equatorial plane ($\theta=\pi/2$, $\dot{\theta}=0$) wlog, so the second element is simply $\lambda_{(2)} = \frac{1}{r} \partial_\theta$. The first element is
\begin{align}
\lambda_{(1)}^t &= (r \dot{r} \cos\psi - E L \sin\psi) \left(1-\frac{2M}{r}\right)^{-1} (r^2 + L^2)^{-1/2} \\
\lambda_{(1)}^r &= (E r \cos\psi - \dot{r} L \sin\psi) (r^2 + L^2)^{-1/2} \\
\lambda_{(1)}^\theta &= 0 \\
\lambda_{(1)}^\phi &= - \frac{(r^2 + L^2)^{1/2}}{r^2} \sin\psi
\end{align}
The third element is
\begin{align}
\lambda_{(3)}^t &= (r \dot{r} \sin\psi + E L \cos\psi) \left(1-\frac{2M}{r}\right)^{-1} (r^2 + L^2)^{-1/2} \\
\lambda_{(3)}^r &= (E r \sin\psi + \dot{r} L \cos\psi) (r^2 + L^2)^{-1/2} \\
\lambda_{(3)}^\theta &= 0 \\
\lambda_{(3)}^\phi &= \frac{(r^2 + L^2)^{1/2}}{r^2} \cos\psi
\end{align}

%The non-zero components of the tidal tensor are
%\begin{align}
%C_{(1)(1)} &= - \left( 1 - 3 \frac{r^2 + L^2}{r^2} \cos^2\psi \right) \frac{M}{r^3} \\
%C_{(2)(2)} &= - \left( 1 + 3 \frac{L^2}{r^2} \right) \frac{M}{r^3} \\
%C_{(3)(3)} &= - \left( 1 - 3 \frac{r^2 + L^2}{r^2} \sin^2\psi \right) \frac{M}{r^3} \\
%C_{(1)(3)} &= C_{(3)(1)} = 3 \frac{r^2 + L^2}{r^5} M \sin\psi \cos\psi
%\end{align}

The angle $\psi$ is calculated from
\begin{equation}
\dot{\psi} = \frac{EL}{r^2 + L^2}
\end{equation}
We take $\psi = 0$ at $r = r_t$, since we chose the point of disruption to be the reference point for our FNCs. This simplifies the above expressions at the tidal radius; in particular, $\lambda_{(1)}^\phi$ vanishes. 
%In particular, $\lambda_{(1)}^\phi$ vanishes, and the tidal tensor becomes diagonal, $C_{(i)(j)} = (M/r^3) \operatorname{diag} \left( (2r^2 + 3L^2)/r^2 , -(1 + 3L^2/r^2) , -1 \right)$.

%%%%%%%%%%%%%%%%%%%%%%%%%%%%%%%%%%%%%%%%%%%%%%%%%%

% Don't change these lines
\bsp	% typesetting comment
\label{lastpage}
\end{document}